\documentclass[preprint,authoryear]{elsarticle}

\usepackage{amsmath}%
\usepackage{amsfonts}%
\usepackage{amssymb}%
\usepackage{graphicx}
\usepackage{booktabs}
\usepackage{url}
\usepackage[all]{xy}
\usepackage{subfig}
\usepackage[noend]{algorithmic}
\usepackage{algorithm}
\usepackage{slashbox}
\usepackage{paralist}
\newenvironment{proof}[1][Proof]{\textbf{#1.} }{\ \rule{0.5em}{0.5em}}
\newtheorem{theorem}{Theorem}

\newcommand{\func}[1]{\ensuremath{\mathit{#1}}}
\newcommand{\var}[1]{\ensuremath{\mathit{#1}}}

\newcommand{\Cluster}[0]{\func{Cluster}}

\newcommand{\Turn}[0]{\func{Turn}}

\newcommand{\heuristic}[1]{\textsf{#1}}

\newcommand{\opt}[3]{\heuristic{#1-opt}$_\text{#2}^\text{#3}$}
\newcommand{\twoopt}[2]{\opt{2}{#1}{#2}}
\newcommand{\optshort}[3]{\heuristic{#1o}$_\text{#2}^\text{#3}$}
\newcommand{\twooptshort}[2]{\optshort{2}{#1}{#2}}
\newcommand{\insertion}[2]{\heuristic{Ins}$_\text{#1}^\text{#2}$}
\newcommand{\Swap}[0]{\heuristic{Swap}}

\newcommand{\FO}[1]{\heuristic{FO}$_\text{#1}$}
\newcommand{\CO}{\heuristic{CO}}

\newcommand{\TourCluster}[2]{\ensuremath{\mathcal{#1}_{#2}}}
\newcommand{\T}[1]{\TourCluster{T}{#1}}

\begin{document}

\title{Efficient Local Search Algorithms for Known and New Neighborhoods for the Generalized Traveling Salesman Problem}

\author[rhul]{D.~Karapetyan\corref{cor1}}
\ead{daniel.karapetyan@gmail.com}
\author[rhul]{G.~Gutin}
\ead{gutin@cs.rhul.ac.uk}

\cortext[cor1]{Corresponding author}
\address[rhul]{Royal Holloway, University of London, Egham, Surrey, TW20 0EX, United Kingdom}


\date{}

\begin{abstract}
The Generalized Traveling Salesman Problem (GTSP) is a well-known combinatorial optimization problem with a host of applications.  It is an extension of the Traveling Salesman Problem (TSP) where the set of cities is partitioned into so-called clusters, and the salesman has to visit every cluster exactly once.

While the GTSP is a very important combinatorial optimization problem and is well studied in many aspects, the local search algorithms used in the literature are mostly basic adaptations of simple TSP heuristics.  Hence, a thorough and deep research of the neighborhoods and local search algorithms specific to the GTSP is required.

We formalize the procedure of adaptation of a TSP neighborhood for the GTSP and classify all other existing and some new GTSP neighborhoods.  For every neighborhood, we provide efficient exploration algorithms that are often significantly faster than the ones known from the literature.  Finally, we compare different local search implementations empirically.

\begin{keyword}
Heuristics, Local Search, Neighborhood, Generalized Traveling Salesman Problem, Combinatorial Optimization.
\end{keyword}
\end{abstract}

\maketitle

\section{Introduction}
\label{sec:intro}

The Generalized Traveling Salesman Problem (GTSP) is an extension of the Traveling Salesman Problem (TSP)\@.  In the GTSP, we are given a set $V$ of $n$ vertices, weights $w(x, y)$ of going from $x \in V$ to $y \in V$ and partition of $V$ into clusters $C_1, C_2, \ldots, C_m$.  A feasible solution, or a \emph{tour}, is a cycle visiting exactly one vertex in every cluster.  The objective is to find the shortest tour.

If the weight matrix is symmetric, i.e., $w(x, y) = w(y, x)$ for any $x, y \in V$, the problem is called \emph{symmetric}.  Otherwise it is an \emph{asymmetric} GTSP.

Observe that the TSP is a special case of the GTSP when $|C_i| = 1$ for each $i$ and, hence, the GTSP is NP-hard.  

The GTSP has a host of applications: warehouse order picking with multiple stock locations, sequencing computer files, postal routing, airport selection and routing for courier planes, and some others, see, e.g., \citep{Fischetti1995,Fischetti1997,Laporte1996,Noon1991} and references therein.

Much attention was paid to solving the GTSP\@.  Several researchers \citep{Ben-Arieh2003,Laporte1999,Noon1993} proposed transformations of a GTSP instance into a TSP instance\@.  At first glance, the idea of transforming a little-studied problem into a well-known one seems to be promising.  However, this approach has a very limited application.  Indeed, it requires exact solutions of the obtained TSP instances because even a near-optimal solution of such TSP may correspond to an infeasible GTSP solution.  At the same time, the produced TSP instances have a rather unusual structure which is hard for the existing TSP solvers.  A more efficient approach to solve the GTSP exactly is the branch-and-bound algorithm designed by~\citet{Fischetti1997}\@.  By using this algorithm, the authors solve several instances of size up to 89 clusters; solving larger instances to optimality is still too hard nowadays.  Two approximation algorithms for special cases of the GTSP were proposed in the literature; alas, the guaranteed solution quality is rather low for the real-world applications, see~\citep{Bontoux2009} and references therein.

In order to obtain good (but not necessarily exact) solutions for larger GTSP instances, one should consider heuristic approach.  Several construction heuristics and local searches were discussed in~\citep{Bontoux2009,GK_GTSP_GA_2008,Hu2008,Renaud1998,Snyder2000} and some others.  A number of metaheuristics were proposed by \citet{Bontoux2009,GK_GTSP_GA_2008,GK_GTSP_GA_2007,Huang2005,Pintea2007,Silberholz2007,Snyder2000,Tasgetiren2007,Yang2008}.  
However, none of these studies provides a review of GTSP neighborhoods or discusses in detail different local search algorithms.  Since most of the solution methods applied to GTSP are somehow based on local search, we believe that a deeper understanding of this subject is of great importance.

In this paper, we define and analyze all known and some new GTSP neighborhoods and the corresponding exploration algorithms.  We consider only the classical local search which guarantees to find a local minimum within a certain neighborhood.  Note that several GTSP neighborhoods were used in \citep{GK_GTSP_GA_2008,GK_GTSP_GA_2007,Snyder2000,Silberholz2007,Tasgetiren2007}, but they were not systematized or analyzed in detail.  We aim to classify all known and new neighborhoods and provide efficient exploration algorithms for all of them.  Note that many of the neighborhoods discussed below are already known from the literature but, because their exploration algorithms were rather slow, some of them were considered practically useless.  Our improvements, of both heuristic and theoretical nature, dramatically speed up the exploration
algorithms, making the corresponding neighborhoods of practical interest.

In our classification, we divide all the GTSP neighborhoods into three classes:
\begin{enumerate}
	\item \emph{Cluster Optimization} neighborhoods consist of solutions which differ from the original one in vertex selection but have the same cluster order.  This class is discussed in Section~\ref{sec:cluster_optimization}.
	\item \emph{TSP-inspired} neighborhoods are GTSP neighborhoods derived from TSP neighborhoods.  Such neighborhoods normally consist of solutions obtained from the original one by some global rearrangements of the cluster order.  The vertex selection within clusters may or may not be preserved in these solutions.  In Section~\ref{sec:ls_adaptation}, we show that there exist several ways to adapt an arbitrary TSP neighborhood to the GTSP and propose a number of ways to make the exploration of these adaptations efficient.
	\item \emph{Fragment Optimization} neighborhoods consist of solutions which are different from the original one in some small tour fragment.  Neighborhoods of this type were not widely used before.  In Section~\ref{sec:fragment_optimization}, we propose two efficient algorithms for exploration of such neighborhoods.
\end{enumerate}

Note that there exists another class of very successful local searches based on the Lin-Kernighan idea \citep{GK_GTSP_LK}, but they are not discusses in this paper because they are not `neighborhood-based.'


\bigskip

In this paper we use the following notation:
\begin{itemize}
	\item $n$ is the number of vertices in the graph.
	\item $m$ is the number of clusters.
	\item $s$ is the maximum cluster size.  Obviously, $\lceil n / m \rceil \le s \le n - m + 1$.
	\item $\gamma$ is the minimum cluster size.  Obviously, $1 \le \gamma \le \lfloor n / m \rfloor$.
	\item $\Cluster(x)$ is the cluster containing vertex $x$.
	\item $w(v_1, v_2)$ is the weight of edge $(v_1, v_2)$.

	\item $w(v_1, v_2, \ldots, v_k) = w(v_1, v_2) + w(v_2, v_3) + \ldots + w(v_{k-1}, v_k)$.
	
	\item $\displaystyle{w_\text{min}(X_1, X_2, \ldots, X_k) = \min_{x_1, x_2, \ldots, x_k} w(x_1, x_2, \ldots, x_k)}$, where $x_i \in X_i$ and $X_i$ is a set of vertices, $i = 1, 2, \ldots, k$.  Function $w_\text{max}(X_1, X_2, \ldots, X_k)$ is defined similarly.
	
	\item $T_i$ denotes the vertex at the $i$th position in tour $T$.  We assume that $T_{i+m} = T_i$.

	\item Tour $T$ is also considered as a set of its edges, i.e., $T = \{ (T_1, T_2),\ (T_2, T_3),\ \ldots,\ (T_{m-1}, T_m),\ (T_m, T_1) \}$.
	
	\item $\Turn(T, x, y)$ denotes the tour obtained from $T$ by reversing the fragment $T_{x+1}$, $T_{x+2}$, \ldots, $T_y$:
	$$
	\Turn(T, x, y) = T_1, \ldots, T_x, \underbrace{T_y, T_{y-1}, \ldots, T_{x+1}}_{Reversed}, T_{y+1}, \ldots, T_m, T_1 \,.
	$$
	Observe that for a symmetric GTSP 
	$$
	\Turn(T, x, y) = T \setminus \{ (T_x, T_{x+1}),\ (T_y, T_{y+1}) \} \cup \{ (T_x, T_y),\ (T_{x+1}, T_{y+1}) \}
	$$
	and, hence, the weight of the obtained tour can be calculated in time $O(1)$:
	\begin{equation}
	\label{eq:turn_delta}
	w(\Turn(T, x, y)) = w(T) - w(T_x, T_{x+1}) - w(T_y, T_{y+1})
	+ w(T_x, T_y) + w(T_{x+1}, T_{y+1}) \,.
	\end{equation}
\end{itemize}

\subsection{Experiments Prerequisites}

Although this paper does not suggest the `best' GTSP local search, as a result of extensive computational experiments, we select the most efficient exploration algorithms and compare different neighborhood variations.  In this section we discuss details of our experimentation techniques.

Our test bed includes several TSP instances taken from TSPLIB~\citep{TSPLIB} and converted to the GTSP by the standard clustering procedure of Fischetti, Salazar, and Toth~\citep{Fischetti1997}; the same approach is widely used in the literature, see, e.g., \citep{GK_GTSP_GA_2008,Silberholz2007,Snyder2000,Tasgetiren2007}.  In particular, we use all the instances with $10 \le m \le 217$ like in~\citep{Bontoux2009,GK_GTSP_GA_2008,Silberholz2007}; in other papers the bounds are more restrictive.  However, to save space, we usually include only every fifth instance in our tables.

Every instance name in the testbed consists of three parts: `$m$ $t$ $n$', where $m$ is the number of clusters, $t$ is the type of the original TSP instance (see~\citep{TSPLIB} for details) and $n$ is the number of vertices.

Observe that the optimal solutions are known only for some instances with at most 89 clusters~\citep{Fischetti1997}.  For the rest of the instances we use the best known solutions, see~\citep{Bontoux2009,GK_GTSP_GA_2008,Silberholz2007}.

In order to generate the starting tour for the local search procedures, we use a simplified Nearest Neighbor~\citep{Noon1988} construction heuristic.  Unlike the algorithm proposed by Noon, our implementation tries only one starting vertex.  According to our experiments, trying every vertex as the starting point significantly slows down the heuristic and almost does not influence the quality of solutions obtained after applying local search.  Note that in what follows, the running time of a local search includes the running time of the construction heuristic.

All the algorithms are implemented in Visual C++; the evaluation platform is based on an Intel~Core~i7 2.67~GHz processor.

\subsection{Local Search Strategy}

Most commonly, one uses the first improvement local search strategy, i.e., applies an improvement as soon as it is found.  Alternatively, one can use the best improvement strategy which first explores the whole neighborhood and then applies the best found improvement.  Note that the first improvement strategy is normally faster while the best improvement strategy gives better solution quality.

We implemented and tested both strategies for most of the algorithms discussed below.  Our experiments show that the difference in solution quality between these two strategies is negligible while the running time of the best improvement is significantly higher. In what follows, we use the first improvement strategy.

\section{Cluster Optimization}
\label{sec:cluster_optimization}

In this section we discuss GTSP neighborhood structures preserving the order of clusters in the tour.  Virtually, the smallest neighborhood of $T$ of this type is 
$$
N_\text{L}(T, i) = \{ (T_1, T_2, \ldots, T_{i-1}, T'_i, T_{i+1}, \ldots, T_m, T_1) :\ T'_i \in \Cluster(T_i) \} \,.
$$
Its size is $|N_\text{L}(T, i)| = |\Cluster(T_i)|$ and it takes $O(s)$ operations to explore it.  One can extend it for two or more clusters: $N_\text{L}(T, I)$, where $I$ is a set of cluster indices to be varied.  The size of such a neighborhood is $|N_\text{L}(T, I)| = \prod_{i \in I} |\Cluster(T_i)|$.

Observe that it takes only $O(|I| s)$ operations to explore $N_\text{L}(T, I)$ if all the clusters selected in $I$ are `independent', i.e., there is no $i$ such that $i \in I$ and $i + 1 \in I$.  If $I = \{ i, i + 1 \}$, the neighborhood $N_\text{L}(T, I)$ changes its structure.  Now it takes $O(s^2)$ operations to explore it.  One may assume that for $I = \{ i, i + 1, \ldots, i + k - 1 \}$ the time complexity of the local search is $O(s^k)$.  Next we will show that, in fact, it takes only $O(ks^2)$ operations to find the best solution in such a neighborhood.

\begin{figure}[ht]
\centerline{
\xymatrix@-1pc@R=5pt@C=45pt{
	&
	&
	&	{\bullet} \ar@{->}[rd] \ar@{->}[rddd] \ar@{->}[rddddd]
\\
	&	{\bullet} \ar@{->}[rd] \ar@{->}[rddd]
	&
	&
	& {\bullet} \ar@{->}[rdd]
\\
	&
	&	{\bullet} \ar@{->}[ruu] \ar@{->}[r] \ar@{->}[rdd] \ar@{->}[rdddd]
	& {\bullet} \ar@{->}[ru] \ar@{->}[rd] \ar@{->}[rddd]
\\
		{\bullet} \ar@{->}[ruu] \ar@{->}[r] \ar@{->}[rdd]
	& {\bullet} \ar@{->}[ru] \ar@{->}[rd]
	&
	&
	& {\bullet} \ar@{->}[r]
	& {\bullet}
\\
	&
	&	{\bullet} \ar@{->}[ruuuu] \ar@{->}[ruu] \ar@{->}[r] \ar@{->}[rdd]
	& {\bullet} \ar@{->}[ruuu] \ar@{->}[ru] \ar@{->}[rd]
\\
	&	{\bullet} \ar@{->}[ruuu] \ar@{->}[ru]
	&
	&
	& {\bullet} \ar@{->}[ruu]
\\
	&
	&
	&	{\bullet} \ar@{->}[ruuuuu] \ar@{->}[ruuu] \ar@{->}[ru]
\\
		{T_{i-1}}
	& {\T{i}}
	& {\T{i + 1}}
	& {\ldots}
	& {\T{i + k - 1}}
	& {T_{i + k}}
}
}
\caption{In order to get the best tour in $N_\text{L}(T, \{ i, i + 1, \ldots, i + k - 1 \})$, construct a layered network as shown here (all the weights in this network correspond to the original weights in the GTSP instance) and find the shortest path from $T_{i - 1}$ to $T_{i + k}$.}
\label{fig:co_local}
\end{figure}
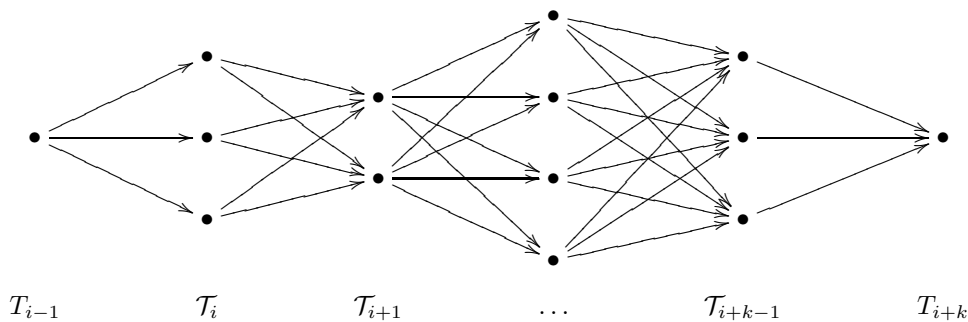

Let $(T_1, T_2, \ldots, T_m, T_1)$ be a tour and $I = \{ i, i + 1, \ldots, i + k - 1 \}$, where $k < m$.  Let $\T{j} = \Cluster(T_j)$.  Construct a layered network as shown in Figure~\ref{fig:co_local}.
Find the shortest path from $T_{i - 1}$ to $T_{i + k}$ in this network and update the vertices in the tour accordingly.  This will yield the shortest tour $T' \in N_\text{L}(T, \{ i, i + 1, \ldots, i + k - 1 \})$, and the time complexity of this algorithm is $O(ks^2)$.

Consider the case where $k = m$.  This is the largest neighborhood of this type and we denote it $N_\text{CO}(T) = N_\text{L}(T, \{1, 2, \ldots, m\})$.  Since $N_\text{CO}(T)$ does not fix any vertices, it is now impossible to use straightforwardly the optimization technique shown above.  However, the problem of finding the shortest tour $T' \in N_\text{CO}(T)$ can be brought to several problems of finding the shortest tour in $N_\text{L}(T, \{2, \ldots, m\})$.  For every $v \in \Cluster(T_1)$ find the shortest tour $T'^v \in N_\text{L}(T^v, \{ 2, 3, \ldots, m \})$, where $T^v = (v, T_2, T_3, \ldots, T_m)$.  The shortest tour among $T'^v$ is the shortest tour $T' \in N_\text{CO}(T)$.  The procedure takes $O(m s^3)$ operations.  In what follows, we call this algorithm \emph{Cluster Optimization} (\CO{}).

\CO{} was introduced by 
\citet{Fischetti1997} (for detailed description see also~\citep{Fischetti2002}) and used in~\citep{GK_GTSP_GA_2008,GK_GTSP_GA_2007,Hu2008,Pintea2007,Renaud1998} and others.

A formal implementation of \CO{} is presented in Algorithm~\ref{alg:co}.
\begin{algorithm}[ht]
\caption{Cluster Optimization.  Basic implementation.}
\label{alg:co}
\begin{algorithmic}
\REQUIRE Tour $T = (T_1, T_2, \ldots, T_m)$.
\STATE Let $\T{i} = \Cluster(T_i)$ for every $i$.
\FORALL {$v \in \T{1}$ and $r \in \T{2}$}
	\STATE Initialize the shortest path from $v$ to $r$: $p_{v, r} \gets (v, r)$.
\ENDFOR
\FOR {$i \gets 3, 4, \ldots, m$}
	\FORALL {$v \in \T{1}$ and $r \in \T{i}$}
		\STATE Set $p_{v, r} \gets p_{v, u} \cup (u, r)$, where $u \in \T{i-1}$ is selected to minimize $w\big(p_{v, u} \cup (u, r)\big)$.
	\ENDFOR
\ENDFOR
\RETURN $p_{v, r} \cup (r, v)$, where $v \in \T{1}$ and $r \in \T{m}$ are selected to minimize $w\big(p_{v, r} \cup (r, v)\big)$.
\end{algorithmic}
\end{algorithm}
Note that $N_\text{CO}(T') = N_\text{CO}(T)$ for any $T' \in N_\text{CO}(T)$ and, thus, unlike usual local search procedures, \CO{} does not need to be run several times to get the local minimum.

\subsection{Cluster Optimization Refinements}
\label{sec:co_refinements}

In this section we discuss several improvements that can noticeably reduce the running time of \CO{}.

\subsubsection{First Cluster Selection}

Observe (see Algorithm~\ref{alg:co}) that the time complexity of \CO{} grows linearly with the size of cluster $\mathcal{T}_1$.  Thus, before applying \CO{}, we rotate the solution such that $|\mathcal{T}_1| = \gamma$.  This technique reduces the time complexity of the algorithm to $O(n \gamma s)$, that was widely used in the literature.

Note that $N_\text{CO}(T)$ is a `very large neighborhood' since it is of an exponential size and there exists a polynomial exploration algorithm for it.  Sometimes, neighborhoods of this class are very effective \citep{GK_Some_Theory}.

\subsubsection{First Cluster Reduction}

Since the running time of \CO{} significantly depends on the size $\gamma$ of the smallest cluster, it is worth checking whether we can reduce its size.  Some attempts to reduce the cluster sizes in the GTSP were proposed by \citet{GK_GTSP_Reduction2008}.  The idea was to remove a vertex $r \in R$, where $R$ is a cluster, if for every pair of vertices $v$ and $u$, $\Cluster(v) \neq \Cluster(u) \neq R$, there exists some $r' \in R \setminus \{ r \}$ such that $w(v, r', u) \le w(v, r, u)$.

In our case, the reduction can be significantly more efficient.  Indeed, we do not need to consider all $u$ and $v$.  Let $R = \mathcal{T}_1$.  Then consider only $u \in \mathcal{T}_m$ and $v \in \mathcal{T}_2$.
	
A straightforward reduction algorithm would take $O(s^2 \gamma^2)$ operations.  We propose Algorithm~\ref{alg:reduce_cluster} which reduces the cluster $\mathcal{T}_1$ in $O(s^2 \gamma)$ time.
\begin{algorithm}[ht]
\caption{Reduction of a cluster in a tour.}
\label{alg:reduce_cluster}

\begin{algorithmic}
\REQUIRE Tour $T = (T_1, T_2, \ldots, T_m, T_1)$, where $|\Cluster(T_1)| = \gamma$.
\STATE Let $U = \Cluster(T_m)$, $R = \Cluster(T_1)$ and $V = \Cluster(T_2)$.
\FORALL {$u \in U$ and $v \in V$}
	\STATE Find the shortest distance $l_{u,v} \gets \min_{r \in R} w(u, r, v)$.
	\STATE Find the number $c_{u,v}$ of paths $(u, r, v)$ such that $w(u, r, v) = l_{u,v}$, i.e., $c_{u,v} \gets \big| \{ r :\ r \in R \text{ and } w(u, r, v) = l_{u,v} \} \big|$.
\ENDFOR

\FORALL {$r \in R$}
	\FORALL {$u \in U$ and $v \in V$}
		\IF {$w(u, r, v) = l_{u,v}$ and $c_{u,v} = 1$}
			\STATE Go to the next $r$.
		\ENDIF
	\ENDFOR
	\FORALL {$u \in U$ and $v \in V$}
		\IF {$w(u, r, v) = l_{u,v}$}
			\STATE Update $c_{u,v} \gets c_{u,v} - 1$.
		\ENDIF
	\ENDFOR
	\STATE Remove $r$ from $R$.
\ENDFOR
\end{algorithmic}
\end{algorithm}
One can try to apply this procedure to reduce every cluster but this would likely slow down the \CO{} algorithm.  We apply this reduction only to the smallest cluster $\mathcal{T}_1 = \Cluster(T_1)$ as shown in Algorithm~\ref{alg:reduce_cluster}.  Moreover, we never apply this reduction if $|\T{m}| |\T{2}| \ge n$.  Indeed, in the best case, \CO{} takes only $\Theta(\gamma n)$ operations (consider, e.g., the case when $|\T{2i-1}| = \gamma$ for every $i$) so it is unreasonable to run the reduction if its time complexity is more than $O(\gamma n)$.

Note that this reduction is valid only for a certain cluster order and, hence, the cluster $\T{1}$ must be restored after the run of \CO{}.

\subsubsection{Calculations Order}

The procedure of finding the shortest paths in a layered network can be described as follows.  Assume that the layers of the network are \T{1}, \T{2}, \ldots, \T{m}, $\T{1}'$, where $\T{1}'$ is a copy of \T{1}, and the objective is to find all the shortest $(v, v')$-paths from every $v \in \T{1}$ to its copy $v' \in \T{1}'$.  Observe that removing any layer \T{i}, $1 < i \le m$, and adding edges from every $u \in \T{i-1}$ to every $v \in \T{i+1}$ such that $w(u, v) = \min_{r \in \T{i}} w(u, r, v)$ preserves the lengths of the shortest $(v, v')$-paths.  After repeating this procedure $m - 2$ times, we get exactly three layers \T{1}, \T{2} and $\T{1}'$ such that $\min_{r \in \T{2}} w(v, r, v')$ is the length of the shortest $(v, v')$-path (we assume that the layers are renumbered after every iteration).  This interpretation is exploited in Algorithm~\ref{alg:co_sequential}.
\begin{algorithm}[ht]
\caption{Sequential \CO{} implementation.  This algorithm is equivalent to Algorithm~\ref{alg:co}.}
\label{alg:co_sequential}

\begin{algorithmic}
\REQUIRE Network layers $\T{1}$, $\T{2}$, \ldots, $\T{m}$, $\T{m+1}$, where $\T{m+1} = \T{1}$.
\FOR {$i \gets 1, 2, \ldots, m - 2$}
	\STATE Set $w(u, v) \gets \min_{r \in \T{2}} w(u, r, v)$ for every $u \in \T{1}$ and $v \in \T{3}$.
	\STATE Remove layer $\T{2}$; renumber the layers accordingly.
\ENDFOR
\RETURN $\min_{v \in \T{1},\,r \in \T{2}} w(v, r, v)$.
\end{algorithmic}
\end{algorithm}

Observe that Algorithm~\ref{alg:co_sequential} removes the layers sequentially but this can be done in an arbitrary order.  A generalized dynamic programming implementation of \CO{} can be described as in Algorithm~\ref{alg:co_general}.
\begin{algorithm}[ht]
\caption{A generalized dynamic programming implementation of \CO{}.}
\label{alg:co_general}

\begin{algorithmic}
\REQUIRE Network layers $\T{1}$, $\T{2}$, \ldots, $\T{m}$, $\T{m+1}$, where $\T{m+1} = \T{1}$.
\FOR {$i \gets 1, 2, \ldots, m - 2$}
	\STATE Set $w(u, v) \gets \min_{r \in \T{X_i}} w(u, r, v)$ for every $u \in \T{X_i - 1}$ and $v \in \T{X_i + 1}$.
	\STATE Remove the layer $\T{X_i}$; renumber the layers accordingly.
\ENDFOR
\RETURN $\min_{v \in \T{1},\,r \in \T{2}} w(v, r, v)$.
\end{algorithmic}
\end{algorithm}
Here $X$ is a sequence of $m - 2$ numbers, $1 < X_i \le m - i + 1$.  It defines the algorithm's behavior: on the $i$th iteration the algorithm removes cluster $\T{X_i}$ from the sequence by calculating the shortest paths from $\T{X_i - 1}$ to $\T{X_i + 1}$.  Note that Algorithms~\ref{alg:co_general} and~\ref{alg:co_sequential} coincide when $X = (2, 2, \ldots, 2)$.

Let us count the number of times Algorithm~\ref{alg:co_general} obtains an edge weight (we will call it \emph{weight operation}).  This number adequately reflects the running time of an implementation.

In general, Algorithm~\ref{alg:co_general} requires
\begin{equation}
\label{eq:co_general_complexity}
t_\text{general} = 2 \cdot \left[ |\T{1}| |\T{k}| + \sum_{i = 1}^{m-2} |\T{x_i}| |\T{y_i}| |\T{z_i}| \right] \text{ weight operations,}
\end{equation}
where $x$, $y$ and $z$ are ordered lists and $1 < k \le m$, all derived from $X$ (we had to introduce these indices because of renumbering performed on every iteration of the algorithm).  Note that in (\ref{eq:co_general_complexity}), the expression in brackets is the number of 3-vertex paths considered by Algorithm~\ref{alg:co_general}, and the factor 2 is the number of weight operations per path.  Without loss of generality, let $x_i < y_i < z_i$.

Algorithm~\ref{alg:co_sequential} always removes the second layer in the current sequence of layers, i.e., the number of weight operations required for the sequential algorithm is as follows:
\begin{equation}
\label{eq:co_sequential_compexity}
t_\text{seq} = 2 \cdot \left[ |\T{1}| |\T{m}| + \sum_{i = 1}^{m-2} |\T{1}| |\T{i+1}| |\T{i+2}| \right] \,.
\end{equation}

Consider the following example.  Let $m$ be even, $|\T{2i}| = z > 1$ and $|\T{2i - 1}| = 1$ for every $i = 1, 2, \ldots, m / 2$.  According to (\ref{eq:co_sequential_compexity}), the sequential algorithm performs $2 (m - 1) z$ weight operations.  Consider the general implementation Algorithm~\ref{alg:co_general} with $X = (2, 3, \ldots, \frac{m}{2}, 2, 2, \ldots, 2)$.  It starts from removing all the layers of size $z$ and then acts as the sequential algorithm.  Observe that it requires only $m z + m - 2$ weight operations.  Hence, the asimptotic ratio is:
$$
\lim_{m \to \infty}{\lim_{z \to \infty}{\frac{2 (m - 1) z}{m z + m - 2}}} = \lim_{m \to \infty}{2 \cdot \frac{m - 1}{m}} = 2 \,.
$$
Note that the weight operations ratio between the sequential calculation and the improved one can be significant in practice.  Even for the modest values $m = 7$ and $z = 7$ in this example the ratio is $1.5$.

The natural question that arises is how much it is possible to speed up the sequential algorithm by changing the calculation order.
\begin{theorem}
Let the first layer in a layered network be the smallest one.  Then the sequential implementation of \CO{} (see Algorithm~\ref{alg:co_sequential}) is at most 2 times slower than the optimal dynamic programming algorithm (see Algorithm~\ref{alg:co_general}), and this bound is asymptotically sharp.
\end{theorem}
\proof 
Let \T{1}, \T{2}, \ldots, \T{m}, $\T{m+1} = \T{1}$ be the layers of the network.  Let $2 < k < m$ (see (\ref{eq:co_general_complexity})).  For every $j = 1, 2, \ldots, m$, equation (\ref{eq:co_general_complexity}) contains a term $|\T{x_i}| |\T{y_i}| |\T{z_i}|$ such that either $x_i = j$ and $y_i = j + 1$ or $y_i = j$ and $z_i = j + 1$.  Indeed, it is impossible to calculate the shortest paths in a layered network without consideration of weights between every pair of consequent layers.  Note that $|\T{x_i}| |\T{y_i}| |\T{z_i}| \ge \gamma |\T{j}| |\T{j+1}|$ if $|\T{x_i}| |\T{y_i}| |\T{z_i}|$ contains $|\T{j}| |\T{j+1}|$.  Observe also that a term $|\T{x_i}| |\T{y_i}| |\T{z_i}|$ may contain both $|\T{j}| |\T{j+1}|$ and $|\T{j+1}| |\T{j+2}|$.  Based on this, we can provide the following lower bound:
\begin{equation}
\label{eq:co_lower_bound}
2 \sum_{i = 1}^{m-2} |\T{x_i}| |\T{y_i}| |\T{z_i}| \ge \gamma \sum_{i = 1}^m |\T{i}| |\T{i+1}| \,.
\end{equation}
Observe that 
$$
\gamma \sum_{i = 1}^m |\T{i}| |\T{i+1}| = |\T{1}| |\T{m}| |\T{m+1}| + \sum_{i = 1}^{m-1} |\T{1}| |\T{i}| |\T{i+1}| \ge \frac{1}{2} t_\text{seq} \,.
$$
Hence, $t_\text{general} \ge \frac{1}{2} t_\text{seq}$.

If $k = 2$, weights between \T{1} and \T{2} are considered in the last line of Algorithm~\ref{alg:co_general} and, hence, (\ref{eq:co_lower_bound}) must be replaced with
$$
2 \sum_{i = 1}^{m-2} |\T{x_i}| |\T{y_i}| |\T{z_i}| \ge \gamma \sum_{i = 2}^m |\T{i}| |\T{i+1}| \,,
$$
that does not change the outcome.

If $k = m$, (\ref{eq:co_lower_bound}) must be replaced with
$$
2 \sum_{i = 1}^{m-2} |\T{x_i}| |\T{y_i}| |\T{z_i}| \ge \gamma \sum_{i = 1}^{m-1} |\T{i}| |\T{i+1}| \,.
$$
In this case 
$$
t_\text{general} = 2 \left[ |\T{1}| |\T{m}| + \sum_{i = 1}^{m-2} |\T{x_i}| |\T{y_i}| |\T{z_i}| \right] \ge 2 |\T{1}| |\T{m}| + \gamma \sum_{i = 1}^{m-1} |\T{i}| |\T{i+1}| \ge \frac{1}{2} t_\text{seq} \,.
$$

The example before the theorem implies that the bound $t_\text{general} \ge \frac{1}{2} t_\text{seq}$ is asymptotically sharp.
\qed

\bigskip

It is not hard to see that the number of distinct dynamic programming implementations of \CO{} is exponential in $m$, and it is usually impractical to search for the optimal calculations order.  Instead, we propose a simple heuristic that improves the sequential algorithm.  On every iteration, out heuristic looks one step ahead; if the condition
\begin{equation}
\label{eq:co_order_condition}
|\T{1}| |\T{2}| |\T{3}| + |\T{1}| |\T{3}| |\T{4}|	> |\T{2}| |\T{3}| |\T{4}| + |\T{1}| |\T{2}| |\T{4}| \,,
\end{equation}
is satisfied for the current numbering of clusters, then it removes cluster \T{3} before removing \T{2}; otherwise it removes \T{2} and proceeds to the next iteration.  For details see Algorithm~\ref{alg:co_improved_order}.
\begin{algorithm}[ht]
\caption{Cluster Optimization with an improved order of calculations.}
\label{alg:co_improved_order}

\begin{algorithmic}
\REQUIRE Tour $T = (T_1, T_2, \ldots, T_m, T_1)$, where $|\Cluster(T_1)| = \gamma$.
\STATE Let $\mathcal{T}_i = \Cluster(T_i)$ for every $i$.
\FOR {$i \gets 2, 3, \ldots, m - 1$}
	\IF {$i < m - 1$ and $|\T{1}| |\T{i}| |\T{i+1}| + |\T{1}| |\T{i+1}| |\T{i+2}|	> |\T{i}| |\T{i+1}| |\T{i+2}| + |\T{1}| |\T{i}| |\T{i+2}|$}
		\STATE Calculate the shortest paths from $\T{i}$ to $\T{i+2}$.
		\STATE Calculate the shortest paths from $\T{1}$ to $\T{i+2}$.
		\STATE Set the weights between $\T{1}$ and $\T{i+2}$ to the calculated values.
		\STATE Set $i \gets i + 1$.
	\ELSE
		\STATE Calculate the shortest paths from $\T{1}$ to $\T{i+1}$.
		\STATE Set the weights between $\T{1}$ and $\T{i+1}$ to the calculated values.
	\ENDIF
\ENDFOR
\RETURN $\min_{v \in \T{1},\,r \in \T{m}} w(v, r, v)$.
\end{algorithmic}
\end{algorithm}

Note that Algorithms~\ref{alg:co_sequential},~\ref{alg:co_general} and~\ref{alg:co_improved_order} find the shortest cycle weight but not the shortest cycle itself.  It will be shown below that it is usually required to find only the weight of the shortest cycle.  In the rare cases that we need the shortest cycle itself, we use the basic sequential implementation (Algorithm~\ref{alg:co}).

\subsection{Computational Experiments}

In order to check the efficiency of the proposed improvements, we provide the results of computational experiments in Tables~\ref{tab:co_variations_one} and~\ref{tab:co_variations_more}.  
\begin{table}[ht]
\scriptsize
\centering
\subfloat[Instances with $\gamma = 1$.]
{
\label{tab:co_variations_one}
\begin{tabular}{lrrcrr}
\toprule

\multicolumn{3}{l}{Instance}&&\multicolumn{2}{c}{Running time, ms}\\
\cmidrule(){1-3}
\cmidrule(){5-6}
Name&$\gamma$&$s$&&\CO{}$_1$&\CO{}$_2$\\
\cmidrule(){1-6}

12brazil58&1&16&&1.1&0.7\\
20kroa100&1&8&&1.8&1.7\\
26bier127&1&27&&2.8&3.0\\
32u159&1&16&&2.3&2.2\\
41gr202&1&17&&4.4&4.4\\
53pr264&1&12&&5.1&5.3\\
87gr431&1&58&&11.8&12.6\\
107att532&1&20&&12.1&12.1\\
131p654&1&25&&20.9&21.3\\
200dsj1000&1&19&&25.8&25.9\\
\cmidrule(){1-6}

Average&1.0&21.8&&8.8&8.9\\
\bottomrule\end{tabular}
}
\qquad
\subfloat[Instances with $\gamma > 1$.]
{
\label{tab:co_variations_more}
\begin{tabular}{lrrcrrrr}
\toprule

\multicolumn{3}{l}{Instance}&&\multicolumn{4}{c}{Running time, ms}\\
\cmidrule(){1-3}
\cmidrule(){5-8}
Name&$\gamma$&$s$&&\CO{}$_1$&\CO{}$_2$&\CO{}$_3$&\CO{}$_4$\\
\cmidrule(){1-8}

10gr48&2&10&&1.1&1.1&0.8&0.8\\
11eil51&2&7&&1.5&1.3&0.9&0.9\\
20rat99&2&11&&3.0&2.8&1.9&1.9\\
20kroc100&2&13&&2.9&2.8&2.3&2.3\\
20krod100&2&9&&3.7&3.1&2.4&2.3\\
20rd100&2&8&&3.0&2.5&3.3&2.9\\
21lin105&2&12&&3.0&2.3&3.1&2.4\\
22pr107&3&7&&5.2&5.1&2.3&2.2\\
25pr124&2&13&&3.8&3.7&2.2&2.2\\
26ch130&2&10&&4.2&3.9&2.5&2.5\\
29pr144&2&10&&4.5&3.9&2.7&2.6\\
30ch150&2&15&&5.6&5.2&3.3&3.3\\
30kroa150&2&11&&5.5&4.9&5.7&5.0\\
36brg180&2&110&&2.7&2.8&2.8&2.9\\
39rat195&2&9&&7.0&6.4&4.0&4.0\\
45ts225&3&9&&8.1&6.0&8.6&6.4\\
56a280&2&10&&9.6&8.9&5.3&5.6\\
207si1032&2&15&&50.1&46.2&27.3&27.4\\
\cmidrule(){1-8}

Average&2.1&16.1&&6.9&6.3&4.5&4.3\\
\bottomrule

\end{tabular}
}
\caption{Experimental results for the different variations of \CO{}\@.  Note that here and below we show the average values in every table.  These should not be considered as the main performance indicators because sometimes they are too much biased to the results obtained for large instances.  However, large instances are of most interest and, thus, averages, being properly understood, can be helpful in analysing experimental results.}
\label{tab:co_implementations}
\end{table}
Table~\ref{tab:co_variations_one} includes only the instances with $\gamma = 1$ (to save space, every fifth instance is taken) while Table~\ref{tab:co_variations_more} includes all the instances with $\gamma > 1$.

All the implementations \CO{}$_1$, \CO{}$_2$, \CO{}$_3$ and \CO{}$_4$ apply the first improvement, i.e., rotate the tour such that $|\Cluster(T_1)| = \gamma$.  In addition, \CO{}$_2$ and \CO{}$_4$ optimize the calculations order according to Algorithm~\ref{alg:co_improved_order}, and \CO{}$_3$ and \CO{}$_4$ try to reduce the size of the smallest cluster according to Algorithm~\ref{alg:reduce_cluster}.

In spite of the fact that all the instances in the test bed have small $\gamma$ (the largest $\gamma$ in the test bed is 3), the experiments clearly show that the cluster reduction technique is very efficient (see the results for \CO$_3$ and \CO$_4$).  It was able to significantly improve the running times for almost every instance in Table~\ref{tab:co_variations_more} (these implementations are obviously not included in Table~\ref{tab:co_variations_one}).

The optimized calculations order is also beneficial, but not so much.  It is more efficient when $\gamma > 1$ (moreover, if $\gamma = 1$, it often slows down the algorithm).  Indeed, it is easy to show that if $\gamma = 1$ then, in order to meet (\ref{eq:co_order_condition}), either \T{2} or \T{4} should be of size 1.  Hence, if $\gamma = 1$, this improvement can be applied quite rarely and only in some relatively easy cases.

We conclude that the proposed refinements are usually insignificant if $\gamma = 1$ but they are very efficient if $\gamma > 1$.  In what follows, we use a hybrid implementation of \CO{}, see Algorithm~\ref{alg:co_hybrid}.
\begin{algorithm}[ht]
\caption{Hybrid implementation of \CO{}.}
\label{alg:co_hybrid}

\begin{algorithmic}
\REQUIRE Tour $T = (T_1, T_2, \ldots, T_m, T_1)$.
\STATE Rotate the tour $T$ such that $|\Cluster(T_1)| = \gamma$.
\IF {$\gamma > 1$}
	\STATE Reduce cluster \T{1} (see Algorithm~\ref{alg:reduce_cluster}).
\ENDIF
\IF {$\gamma = 1$}
	\STATE Apply sequential implementation of \CO{} (see Algorithm~\ref{alg:co_sequential}).
\ELSE
	\STATE Apply \CO{} with improved calculations order (see Algorithm~\ref{alg:co_improved_order}).
\ENDIF
\RETURN $T$.
\end{algorithmic}
\end{algorithm}

\section{TSP-inspired Neighborhoods}
\label{sec:tsp_neighborhoods}

Since the GTSP is an extension of the TSP, it is natural to use TSP neighborhood adaptations for the GTSP\@.  In this section we discuss different ways to adapt a TSP neighborhood for the GTSP.  These approaches are later applied to the most efficient TSP neighborhoods.  Note that some of these ideas are presented in~\citep{GK_GTSP_LK} but in this study they are generalized, further developed and discussed in detail.

It is worth saying that the adaptation of a TSP neighborhood for the GTSP is not as straightforward as it may seem to be.  Among other approaches, \citet{Renaud1998} propose decomposing GTSP into two problems: solving the TSP instance induced by the given tour to find the cluster order and then applying \CO{} algorithm to it (see Section~\ref{sec:cluster_optimization}).  We will show now that this method is generally poor with regard to solution quality.  Let $N_\text{ind}(T)$ be a set of tours which can be obtained from the tour $T$ by reordering vertices in $T$.  Observe that one has to solve a TSP instance induced by $T$ to find the best tour in $N_\text{ind}(T)$.  Let $N_\text{CO}(T)$ be the neighborhood of the \CO{} local search (see Section~\ref{sec:cluster_optimization}).  

The following theorem shows that decomposing the GTSP into two problems (iteratively search in $N_\text{ind}(T)$ and then search in $N_\text{CO}(T)$) does not guarantee any solution quality.  For a proof, see \citet{GK_GTSP_LK}.

\begin{theorem}
\label{th:tsp_co_local_minimum}
The best tour among $N_\text{CO}(T) \cup N_\text{ind}(T)$ can be a longest GTSP tour different from a shortest one.
\end{theorem}

\subsection{TSP Neighborhoods}

In order to continue this discussion, let us briefly list the most well-known TSP neighborhoods.  Here we assume that $m$ is the number of vertices in the TSP instance.

\begin{description}
	\item[$k$-opt] is the most general TSP neighborhood.  It includes all the tours that are different from the given one in at most $k$ edges.  Obviously any tour can be obtained from a given one by an $m$-opt move.
	
	\item[Insertion] neighborhood includes all the tours that can be obtained from the given one by removing a vertex and inserting it at
some other position.  It can be viewed as a special case of 3-opt.

	\item[Swap (also known as Exchange)] neighborhood includes all the tours that can be obtained from the given one by swapping two vertices.  It can be viewed as a special case of 4-opt.
		
	\item[Lin-Kernighan] is a sophisticated heuristic exploring some areas of $k$-opt without fixing the value of $k$.  It does not have any certain neighborhood and, thus, is not considered in this paper.
\end{description}

For more information on these and some other TSP local searches, see, e.g., \citep{Johnson2002,Johnson2002a}.

\subsection{Adaptation of TSP local search for GTSP}
\label{sec:ls_adaptation}

A typical local search with a neighborhood $N(T)$ is shown in Algorithm~\ref{alg:typical_ls}.
\begin{algorithm}[ht]
\caption{Typical local search with neighborhood $N(T)$.}
\label{alg:typical_ls}
\begin{algorithmic}
\REQUIRE Solution $T$.
\FORALL {$T' \in N(T)$}
	\IF {$w(T') < w(T)$}
		\STATE $T \gets T'$.
		\STATE Rerun the for loop again.
	\ENDIF
\ENDFOR
\RETURN $T$.
\end{algorithmic}
\end{algorithm}
Let $N_1(T) \subseteq N_\text{ind}(T)$ be a neighborhood of some TSP local search $\func{LS}_1(T)$.  Let $N_2(T) \subseteq N_\text{CO}(T)$ be a neighborhood of the Cluster Optimization class and $\func{LS}_2(T)$ an exploration algorithm for it.  Then one can think of the following two ways to combine these local searches in one GTSP local search:
\begin{enumerate}[(i)]
	\item \label{item:coinsidetsp} Enumerate all candidates $T' \in N_1(T)$.  For every candidate $T'$ find $T'' \gets \func{LS}_2(T')$ to optimize it in $N_2(T')$.  If $w(T'') < w(T)$, replace $T$ with $T''$ and continue.
	\item \label{item:tspinsideco} Enumerate all candidates $T' \in N_2(T)$.  For every candidate $T'$ find $T'' \gets \func{LS}_1(T')$ to optimize it in $N_1(T')$.  If $w(T'') < w(T)$, replace $T$ with $T''$ and continue.
\end{enumerate}

Observe that the neighborhood $N_1(T)$ is normally much harder to explore than the cluster optimization neighborhood $N_2(T)$.  Consider, e.g., $N_1(T) = N_\text{ind}(T)$ and $N_2(T) = N_\text{CO}(T)$.  Then both options yield an optimal GTSP solution but Option~(\ref{item:coinsidetsp}) requires only $O(n \gamma s \cdot (m - 1)!)$ operations while Option~(\ref{item:tspinsideco}) requires $O(s^m \cdot (m - 1)!)$ operations.

Moreover, many practical applications of the GTSP have some localization of clusters, i.e., typically, $|w(x, y_1) - w(x, y_2)| \ll w(x, y_1)$ if $\Cluster(y_1) = \Cluster(y_2) \neq \Cluster(x)$.  Hence, the dependency of the $N_2(T)$ landscape on the cluster order is higher than the dependency of the $N_1(T)$ landscape on the vertex selection and, thus, Option~(\ref{item:coinsidetsp}) is preferable.

Option~(\ref{item:tspinsideco}) was used by~\citet{Hu2008}.  Note that using $N_2(T) = N_\text{CO}(T)$ would lead to a non-polynomial algorithm; the cluster optimization neighborhood $N_2(T)$ they use includes only the tours which differ from $T$ in exactly one vertex.  For every $T' \in N_2(T)$, the Chained Lin-Kernighan heuristic is applied.  This results in $n$ runs of Chained Lin-Kernighan which makes the algorithm unreasonably slow while the vertex selection is given a very little freedom.

\bigskip

Option~(\ref{item:coinsidetsp}) may be improved as shown in Algorithm~\ref{alg:co_inside_tsp_improved}.
\begin{algorithm}[ht]
\caption{Improved adaptation of a TSP neighborhood for the GTSP according to Option~(\ref{item:coinsidetsp}).}
\label{alg:co_inside_tsp_improved}
\begin{algorithmic}
\REQUIRE Tour $T$.
\FORALL {$T' \in N_1(T)$}
	\STATE $T' \gets \func{QuickImprove}(T')$.
	\IF {$w(T') < w(T)$}
		\STATE $T \gets \func{SlowImprove}(T')$.
		\STATE Rerun the for loop again.
	\ENDIF
\ENDFOR
\RETURN $T$.
\end{algorithmic}
\end{algorithm}
Here $\func{QuickImprove}(T)$ and $\func{SlowImprove}(T)$ are some tour improvement heuristics of the Cluster Optimization class.  Formally, these heuristics should meet the following requirements: 
\begin{itemize}
	\item $\func{QuickImprove}(T), \func{SlowImprove}(T) \in N_\text{CO}(T)$ for any tour $T$;
	\item $w(\func{QuickImprove}(T)) \le w(T)$ and $w(\func{SlowImprove}(T)) \le w(T)$ for any tour $T$.
\end{itemize}
\func{QuickImprove} is applied to every candidate $T'$ before its evaluation.  \func{SlowImprove} is only applied to successful candidates in order to further improve them.  One can think of the following implementations of \func{QuickImprove} and \func{SlowImprove}: 
\begin{itemize}
	\item Trivial $I(T)$ which leaves the solution without any change: $I(T) = T$.
	\item Local cluster optimization $L(T) = L(T, I)$, see Section~\ref{sec:cluster_optimization}.  It updates vertices only within clusters $C_i$, $i \in I$, affected by the latest solution change.  E.g., if a tour $(x_1, x_2, x_3, x_4, x_1)$ was changed to $(x_1, x_3, x_2, x_4, x_1)$, we can use $L(T, \{ 2, 3 \})$ which will yield the best solution among $(x_1, x'_3, x'_2, x_4, x_1)$, where $x'_2 \in \Cluster(x_2)$ and $x'_3 \in \Cluster(x_3)$.  The time complexity of $L(T)$ is $O(|I| s)$ or $O(|I| s^2)$, depending on the affected clusters.
	\item Global cluster optimization $\mathit{CO}(T)$ which applies the \CO{} algorithm to the given solution.  The time complexity of \CO{} is $O(n \gamma s)$.
\end{itemize}

There are five meaningful combinations of \func{QuickImprove} and \func{SlowImprove}:
\begin{description}
	\item[Basic] $\func{QuickImprove}(T) = I(T)$ and $\func{SlowImprove}(T) = I(T)$.  This actually yields the original TSP local search applied to the TSP instance induced by the GTSP tour $T$.

	\item[Basic with CO] $\func{QuickImprove}(T) = I(T)$ and $\func{SlowImprove}(T) = \func{CO}(T)$, i.e., the algorithm explores the original TSP neighborhood but every time an improvement $T'$ is found, it is optimized in $N_\text{CO}(T')$.  One can also consider $\func{SlowImprove}(T) = L(T)$, but such adaptation has no practical interest.  Indeed, \func{SlowImprove} is used quite rarely and so its influence on the total running time is negligible.  At the same time, $\func{CO}(T)$ is much more powerful than $L(T)$ with respect to solution quality.

	\item[Local] $\func{QuickImprove}(T) = L(T)$ and $\func{SlowImprove}(T) = I(T)$, i.e., every candidate $T' \in N_1(T)$ is improved locally before it is compared to the original solution.

	\item[Local with CO] $\func{QuickImprove}(T) = L(T)$ and $\func{SlowImprove}(T) = \func{CO}(T)$, which is the same as \emph{Local} but in addition it optimizes every improvement $T'$ globally in $N_\text{CO}(T')$.

	\item[Global] $\func{QuickImprove}(T) = \func{CO}(T)$ and $\func{SlowImprove}(T) = I(T)$, i.e., every candidate $T' \in N_1(T)$ is optimized globally in $N_\text{CO}(T')$ before it is compared to the original solution $T$.
\end{description}

For a TSP local search \func{LS} we use $\func{LS}_\text{B}$, $\func{LS}_\text{B}^\text{co}$, $\func{LS}_\text{L}$, $\func{LS}_\text{L}^\text{co}$ and $\func{LS}_\text{G}$ to denote the Basic, Basic with CO, Local, Local with CO and Global adaptations of \func{LS}, respectively.

Some of these adaptations were applied in the literature.  For example, the heuristics G2 and G3~\citep{Renaud1998} are actually Global adaptations of 2-opt and 3-opt TSP heuristics, respectively.  An enhanced implementation of the Global 2-opt adaptation is proposed by~\citet{Hu2008}; asymptotically, it is faster than the naive implementation by factor 3.  Local adaptations of 2-opt and some other neighborhoods were used by~\citet{Fischetti1997,GK_GTSP_GA_2008,Silberholz2007,Snyder2000,Tasgetiren2007}.  Some Basic adaptations were used by~\citet{Bontoux2009,GK_GTSP_GA_2008,Silberholz2007,Snyder2000}.

\subsection{Global Adaptation}
\label{sec:global}

The most powerful adaptation of a TSP local search for the GTSP is the Global adaptation.  It applies \CO{} to every candidate tour before it is evaluated.  In other words, if $N_1(T) \subseteq N_\text{ind}(T)$ is the original TSP neighborhood, then the adapted neighborhood $N(T)$ is as follows:
$$
N(T) = \bigcup_{T' \in N_1(T)} N_\text{CO}(T') \,.
$$
Observe that the Global adaptation turns a polynomial size TSP neighborhood into a very large neighborhood, i.e., into a neighborhood of the exponential size that can be explored in polynomial time.  Indeed, $N_\text{CO}(T_1) \cap N_\text{CO}(T_2) = \varnothing$ if the tours $T_1$ and $T_2$ have different cluster order.  Hence, the size of $N(T)$ is exactly 
$$
|N(T)| = |N_1(T)| \cdot \prod_{i=1}^m{|C_i|} \in O(|N_1(T)| \cdot s^m) \,,
$$
while it takes only $O(|N_1(T)| \cdot \gamma s n)$ operations to explore it.  This approach was applied by \citet{Renaud1998} and it was slightly improved by \citet{Hu2008}.

We propose a new technique that can further speed up the Global adaptation.  In particular, it is $m \gamma / s$ times faster than a straightforward adaptation described above.  It was first applied in \citep{GK_GTSP_LK} for the Lin-Kernighan heuristic.  In this paper we generalize this approach and also provide some additional improvements.

The main idea of our technique is to generate candidates $T' \in N_1(T)$ in a certain order such that previously calculated shortest paths could be reused.  Observe that any TSP local search is a special case of $k$-opt.  Indeed, any transformation of a TSP tour may be represented as a $k$-opt move, subject to a sufficiently large value of $k$.

Let $\func{k\text{-opt}}(T, \alpha, \beta)$ be a tour obtained from $T$ by removing edges $\alpha$ and adding edges $\beta$, where $\alpha$ and $\beta$ are edge sets, $|\alpha| = |\beta| = k$.  We need to group all the candidates $T' \in N_1(T)$ into $g$ groups, each group meeting the following requirements:
\begin{itemize}
	\item Let $T^1$, $T^2$, \ldots, $T^l$ be a group of candidates and $T^i = \func{k\text{-opt}}(T, \alpha^i, \beta^i)$.  Without loss of generality, we may assume that $k = \text{const}$ for the whole group of candidates.
	
	\item Let $\alpha = \bigcap_{i=1}^l {\alpha^i}$ and let $\alpha'^i = \alpha^i \setminus \alpha$.  Similarly, $\beta = \bigcap_{i=1}^l {\beta^i}$.
	
	\item Let $Q = (T \setminus \alpha) \cup \beta$, i.e., $Q$ is a set of paths and/or cycles produced from $T$ by removing the edges $\alpha$ and adding the edges $\beta$.
	
	\item Removing the edges $\alpha'^i$ from $Q$ yields a number of paths, let us say $P^i_1, P^i_2, \ldots, P^i_{k - |\beta|}$.  Our requirement for each group is that every of these paths has at least one fixed end:
\begin{align*}
\func{beginning}(P^i_x) &= \func{beginning}(P^j_x) \text{ for every $i, j \in \{ 1, 2, \ldots, l \}$, or} \\
\func{end}(P^i_x) &= \func{end}(P^j_x) \text{ for every $i, j \in \{ 1, 2, \ldots, l \}$}
\end{align*}
for every $x = 1, 2, \ldots, {k - |\beta|}$.

	\item In order to achieve an $m \gamma / s$ times speed up, the number $g$ of groups must be $g \in O(\frac{|N_1(T)|}{m})$, and the number of edges in every $\alpha^i$ must be fixed: $k - |\alpha| \in O(1)$.
\end{itemize}

If the above requirements are satisfied, the Global adaptation may be implemented as in Algorithm~\ref{alg:global_adaptation}.
\begin{algorithm}[ht]
\caption{General implementation of the Global adaptation of a TSP local search.}
\label{alg:global_adaptation}

\begin{algorithmic}
\REQUIRE Tour $T$ optimal in $N_\text{CO}(T)$, i.e., $T = \CO(T)$.
\REQUIRE A group of candidates $T^1, T^2, \ldots, T^l$ such that $T^i = \func{k\text{-opt}}(T, \alpha^i, \beta^i)$ for $i = 1, 2, \ldots, l$.
\STATE Let $\alpha \gets \bigcap_{i=1}^l {\alpha^i}$ and $\beta \gets \bigcap_{i=1}^l {\beta^i}$. Let $\alpha'^i \gets \alpha^i \setminus \alpha$ for $i = 1, 2, \ldots, l$.
\STATE Let $Q \gets (T \setminus \alpha) \cup \beta$.  Let $Q \setminus \alpha^i = \{ P^i_1, P^i_2, \ldots, P^i_{k - |\beta|}\}$ for $i = 1, 2, \ldots, l$.  Note that the paths $P^i_j$ have to meet the conditions above, see Section~\ref{sec:global}.
\FOR {$j \gets 1, 2, \ldots, k - |\beta|$}
	\STATE Calculate all the shortest paths through the cluster sequences corresponding to $P^1_j$, $P^2_j$, \ldots, $P^l_j$.
\ENDFOR
\FOR {$i \gets 1, 2, \ldots, l$}
	\STATE Construct a layered network $L$ as follows:
	\begin{compactitem}
		\item Each layer $2j - 1$, $j = 1, 2, \ldots, k - |\beta|$, corresponds to the cluster $\func{beginning}(P^i_j)$;
		\item Each layer $2j$, $j = 1, 2, \ldots, k - |\beta|$, corresponds to the cluster $\func{end}(P^i_j)$;
		\item The weights between layers $2j - 1$ and $2j$ are equal to the shortest paths in $P^i_j$;
		\item The weights between layers $2j$ and $2j + 1$ are equal to the weights between corresponding clusters.
		\item Layer $2(k - |\beta|) + 1$ is a copy of layer 1, and the weights between layers $2(k - |\beta|)$ and $2(k - |\beta|) + 1$ are equal to the weights between corresponding clusters.
	\end{compactitem}

	\STATE Find the shortest cycle $C$ in the layered network $L$ using the \CO{} algorithm.
	
	\IF {$w(C) < w(T)$}
		\STATE Update tour $T$ according to the cycle $C$.
		\STATE Restart the algorithm.
	\ENDIF
\ENDFOR
\RETURN $T$.
\end{algorithmic}
\end{algorithm}
Observe that finding the shortest paths in a series of fragments $P_j^1$, $P_j^2$, \ldots, $P_j^l$ takes only $O(n s^2)$ operations: start from the fixed end of $P_j^i$ and calculate the shortest paths to every vertex in the required direction.  Since the number of fragments $k - |\beta|$ is fixed, finding the shortest paths in all fragments $P_j^i$, $i = 1, 2, \ldots, k - |\beta|$, $j = 1, 2, \ldots, l$, also takes $O(n s^2)$ time.  All the runs of \CO{} take $O(n s^2)$ operations.  Thus, instead of $O(m n \gamma s)$ operations needed for a `naive' implementation to explore a group of $\Theta(m)$ candidates, Algorithm~\ref{alg:global_adaptation} takes $O(n s^2)$ time.

Observe that this algorithm can be used for both symmetric and assymmetric GTSP\@.  Indeed, even if orientation of some path in the candidate tour does not coincide with orientation of this path in the original tour, one can calculate the shortest paths within this fragment in the backward direction.

\subsubsection{Implementation Example}
\label{sec:two_opt_global_adaptation}

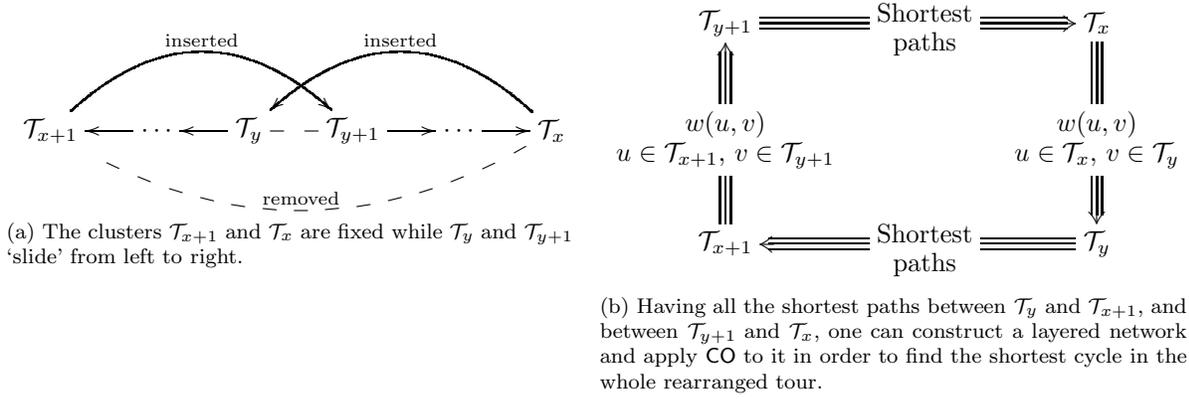
\begin{figure}[ht]
\centering
\subfloat[The clusters $\mathcal{T}_{x+1}$ and $\mathcal{T}_x$ are fixed while $\mathcal{T}_y$ and $\mathcal{T}_{y+1}$ `slide' from left to right.]
{
	\label{fig:two_opt_global_tour}
\xymatrix@R=3em@C=1.8em@L=0.2em{
\\
		*+{\mathcal{T}_{x+1}} \ar@{<-}[r]
	&	*+{\cdots} \ar@{<-}[r]
	&	*+{\mathcal{T}_y} \ar@/^3em/@{<-}[rrr]^{\text{inserted}} \ar@{--}[r]
	&	*+{\mathcal{T}_{y+1}} \ar@/_3em/@{<-}[lll]_{\text{inserted}}
	&	*+{\cdots} \ar@{<-}[l]
	&	*+{\mathcal{T}_x} \ar@{<-}[l] \ar@/^3em/@{--}[lllll]_{\text{removed}}
}
}
\quad
\subfloat[Having all the shortest paths between $\mathcal{T}_y$ and $\mathcal{T}_{x+1}$, and between $\mathcal{T}_{y+1}$ and $\mathcal{T}_x$, one can construct a layered network and apply \CO{} to it in order to find the shortest cycle in the whole rearranged tour.]
{
	\label{fig:two_opt_global_network}
\xymatrix@R=1.5em@C=1em@L=0.5em{
		*+{\mathcal{T}_{y+1}} \ar@3{-}[r]
	&	\txt{Shortest\\paths} \ar@3{->}[r]
	&	*+{\mathcal{T}_x} \ar@3{-}[d]
\\
		\txt{$w(u, v)$\\$u \in \mathcal{T}_{x+1}$, $v \in \mathcal{T}_{y+1}$} \ar@3{->}[u]
	&
	&	\txt{$w(u, v)$\\$u \in \mathcal{T}_x$, $v \in \mathcal{T}_y$} \ar@3{->}[d]
\\
		*+{\mathcal{T}_{x+1}} \ar@3{-}[u]
	&	\txt{Shortest\\paths} \ar@3{->}[l]
	&*+{\mathcal{T}_y} \ar@3{-}[l]
}
}
\caption{Global adaptation of the 2-opt heuristic.}
\label{fig:two_opt_global}
\end{figure}

Let us consider the 2-opt TSP neighborhood and its Global adaptation.  Algorithm~\ref{alg:two_opt_enumerate} enumerates all the candidates in $N_\text{2-opt}(T)$.
\begin{algorithm}[ht]
	\caption{Enumeration of all the candidates in the TSP 2-opt neighborhood.}
	\label{alg:two_opt_enumerate}
\begin{algorithmic}
\REQUIRE Tour $T$.
\FOR {$x \gets 1, 2, \ldots, m - 2$}
	\FOR {$y \gets x + 2, x + 3, \ldots, \min\{m, x + m - 2\}$}
		\STATE List the candidate $\Turn(T, x, y)$ (see Section~\ref{sec:intro}).
	\ENDFOR
\ENDFOR
\end{algorithmic}
\end{algorithm}
Consider a group of candidates $\{ T^1, T^2, 
\ldots, T^l \} \subset N_\text{2-opt}(T)$ such that $T^i = \func{k\text{-opt}}(T, \alpha^i, \beta^i)$ for $i = 1, 2, \ldots, l$, where $\alpha^i = \{ (T_x, T_{x+1}),\, (T_{y(i)}, T_{y(i)+1}) \}$ and $\beta^i = \{ (T_x, T_{y(i)}),\, (T_{x+1}, T_{y(i)+1}) \}$ (see Figure~\ref{fig:two_opt_global_tour}).
We get $\alpha = \{ (T_x, T_{x+1}) \}$ and $\beta = \varnothing$.  Hence, $Q$ is a path obtained from $T$ by removing the edge $(T_x, T_{x+1})$.  Further removing the edge $\alpha'^i = \{ (T_{y(i)}, T_{y(i)+1}) \}$ splits $Q$ into two paths $(T_{x+1}, \ldots, T_{y(i)})$ and $(T_{y(i)+1}, \ldots, T_x)$.  Observe that $(T_{x+1}, \ldots, T_{y(i)})$ has a fixed beginning, and $(T_{y(i)+1}, \ldots, T_x)$ has a fixed end.  Observe also that the number of candidate groups is $\Theta(m)$ while the total number of TSP candidates is $\Theta(m^2)$, and, hence, $g \in O( \frac{|N_\text{2-opt}(T)|}{m} )$. 

Algorithm~\ref{alg:two_opt_global} explores the neighborhood $N_\text{2-opt}(T)$ for some fixed $x$.
\begin{algorithm}[ht]
	\caption{Global adaptation of 2-opt.}
	\label{alg:two_opt_global}
\begin{algorithmic}
\REQUIRE Tour $T$.
\STATE Let $\mathcal{T}_i \gets \Cluster(T_i)$.
\FOR {$x \gets 1, 2, \ldots, m - 2$}
	\STATE Calculate the shortest paths along the tour $T$ from every vertex in $\mathcal{T}_y$ to every vertex in $\mathcal{T}_{x+1}$ and from every vertex in $\mathcal{T}_{y+1}$ to every vertex in $\mathcal{T}_x$ for every $y = x + 2, x + 3, \ldots, \min\{m, x + m - 2\}$.
	\FOR {$y \gets x + 2, x + 3, \ldots, \min\{m, x + m - 2\}$}
		\STATE Construct a layered network $L$ as in Figure~\ref{fig:two_opt_global_network}.
		\STATE Apply \CO{} to $L$ to get the shortest cycle $C$.
		\IF {$w(C) < w(T)$}
			\STATE Replace $T$ with $C$.
			\STATE Restart the whole algorithm.
		\ENDIF
	\ENDFOR
\ENDFOR
\end{algorithmic}
\end{algorithm}
Compare the time complexity of the naive exploration of $N_\text{2-opt}(T)$, which is $O(m^2 n \gamma s)$, with our adaptation, which takes only $O(m n s^2)$ operations.  If $s / \gamma \ll m$, which is a very natural assumption, our implementation is significantly faster than the naive one.

\subsection{Global Adaptation Refinements}
\label{sec:global_refinements}

Observe that the above proposed implementation of the global adaptation consists of 
\begin{inparaenum}[(a)]
	\item \label{item:calc_shortest_paths} calculating the shortest paths through tour fragments, and
	\item \label{item:calc_shortest_cycles} calculating the shortest cycles.
\end{inparaenum}
Both parts are time consuming; for example, in \twoopt{G}{}, each (\ref{item:calc_shortest_paths}) and (\ref{item:calc_shortest_cycles}) takes $O(m n s^2)$ operations.  In Section~\ref{sec:lower_bound} we try to predict if a candidate can improve current solution without running \CO{}\@.  This only or almost only affects part~(\ref{item:calc_shortest_cycles}).  To improve (\ref{item:calc_shortest_paths}), in Sections~\ref{sec:supporting_cluster} and~\ref{sec:multiple_supporting_clusters} we propose an approach that dramatically reduces the number of shortest paths to be calculated.  It also saves time on part~(\ref{item:calc_shortest_cycles}) by selecting smaller clusters for the layers in networks $L$.

\subsubsection{Lower Bound}
\label{sec:lower_bound}

In the proposed adaptation, we calculate the shortest cycle on every iteration.  Having a lower bound for the shortest cycle, one could omit some of these calculations.

Assume that the rearranged tour $T$ consists of $k$ cluster sequences $P^1$, $P^2$, \ldots, $P^k$ such that $\func{end}(P^i)$ is connected to $\func{beginning}(P^{i+1})$ and $\func{end}(P^k)$ is connected to $\func{beginning}(P^1)$, where $\func{beginning}(P^i)$ ($\func{end}(P^i)$) is the first (the last) cluster in $P^i$.  Let $p^i$ be the weight of the shortest path through the cluster sequence $P^i$.  Then the following is a lower bound for the shortest cycle in this sequence of clusters:
$$
w(\func{CO}(T)) \ge \sum_{i=1}^k \left[p^i + w_\text{min}\left(\func{beginning}(P^i),\, \func{end}(P^{i+1})\right)\right] \,,
$$
where $P^{k+1} = P^1$.  Recall that $w_\text{min}(X, Y)$ is the weight of the shortest edge from cluster $X$ to cluster $Y$.

It would take too much time to calculate the shortest paths $p^i$ on every iteration.  Instead, we propose a lower bound for $p^i$ according to Theorem~\ref{th:shortest_path_lower_bound}.

\begin{theorem}
\label{th:shortest_path_lower_bound}
For the shortest path from an arbitrary vertex in $\mathcal{T}_a$ to an arbitrary vertex in $\mathcal{T}_b$ in a layered network $\mathcal{T}_1 \cup \mathcal{T}_2 \cup \ldots \cup \mathcal{T}_m$ we have:
\begin{multline}
\label{eq:shortest_path_lower_bound}
w_\text{min}(\mathcal{T}_a, \mathcal{T}_{a+1}, \ldots, \mathcal{T}_b) \ge w(T_a, T_{a+1}, \ldots, T_b)\\
- w_\text{max}(T_a, \mathcal{T}_{a+1}) - w_\text{max}(\mathcal{T}_{b-1}, T_b) + w_\text{min}(\mathcal{T}_a, \mathcal{T}_{a+1}) + w_\text{min}(\mathcal{T}_{b-1}, \mathcal{T}_b) \,,
\end{multline}
where $(T_1, T_2, \ldots, T_m, T_1)$ is the shortest cycle through all the layers of the network.
\end{theorem}
\proof
Observe that $(T_a, T_{a+1}, \ldots, T_b)$ is the shortest path from $T_a$ to $T_b$ through the layers $\mathcal{T}_{a+1}$, $\mathcal{T}_{a+2}$, \ldots, $\mathcal{T}_{b-1}$.  Indeed, if there was a shorter path, the shortest cycle $(T_1, T_2, \ldots, T_m, T_1)$ could be improved.

Assume that there exists some path $(T'_a, T'_{a+1}, \ldots, T'_b)$, $T'_i \in \T{i}$, shorter than the lower bound provided in~(\ref{eq:shortest_path_lower_bound}):
\begin{multline}
\label{eq:assumption}
w(T'_a, T'_{a+1}, \ldots, T'_b) < w(T_a, T_{a+1}, \ldots, T_b)\\
- w_\text{max}(T_a, \mathcal{T}_{a+1}) - w_\text{max}(\mathcal{T}_{b-1}, T_b) + w_\text{min}(\mathcal{T}_a, \mathcal{T}_{a+1}) + w_\text{min}(\mathcal{T}_{b-1}, \mathcal{T}_b) \,.
\end{multline}
Observe that 
\begin{align}
\label{eq:left_inequality}
w(T_a, T'_{a+1}) - w_\text{max}(T_a, \mathcal{T}_{a+1}) &\le w(T'_a, T'_{a+1}) - w_\text{min}(\mathcal{T}_a, \mathcal{T}_{a+1}) \text{ and} \\
\label{eq:right_inequality}
w(T'_{b-1}, T_b) - w_\text{max}(\mathcal{T}_{b-1}, T_b) &\le w(T'_{b-1}, T'_b) - w_\text{min}(\mathcal{T}_{b-1}, \mathcal{T}_b)
\end{align}
because the left-hand sides of (\ref{eq:left_inequality}) and (\ref{eq:right_inequality}) are non-positive and the right-hand sides are non-negative.  We have
$
w(T'_a, T'_{a+1}, \ldots, T'_b)
= w(T'_a, T'_{a+1}) + w(T'_{a+1}, T'_{a+2}, \ldots, T'_{b-1}) + w(T'_{b-1}, T'_b)
$. 
By substitution of lower bound for $w(T'_a, T'_{a+1})$ and $w(T'_{b-1}, T'_b)$ obtained from (\ref{eq:left_inequality}) and (\ref{eq:right_inequality}), respectively, to (\ref{eq:assumption}) we get:
\begin{multline*}
w(T_a, T'_{a+1}) - w_\text{max}(T_a, \mathcal{T}_{a+1}) + w_\text{min}(\mathcal{T}_a, \mathcal{T}_{a+1}) + w(T'_{a+1}, T'_{a+2}, \ldots, T'_{b-1})\\
+ w(T'_{b-1}, T_b) - w_\text{max}(\mathcal{T}_{b-1}, T_b) + w_\text{min}(\mathcal{T}_{b-1}, \mathcal{T}_b)\\
< w(T_a, T_{a+1}, \ldots, T_b)
- w_\text{max}(T_a, \mathcal{T}_{a+1}) - w_\text{max}(\mathcal{T}_{b-1}, T_b) + w_\text{min}(\mathcal{T}_a, \mathcal{T}_{a+1}) + w_\text{min}(\mathcal{T}_{b-1}, \mathcal{T}_b) \,.
\end{multline*}
From that we have
\begin{align*}
w(T_a, T'_{a+1}) + w(T'_{a+1}, T'_{a+2}, \ldots, T'_{b-1}) + w(T'_{b-1}, T_b) &< w(T_a, T_{a+1}, \ldots, T_b) \text{ or} \\
w(T_a, T'_{a+1}, T'_{a+2}, \ldots, T'_{b-1}, T_b) &< w(T_a, T_{a+1}, \ldots, T_b) \,.
\end{align*}
Hence, the path $(T_a, T'_{a+1}, T'_{a+2}, \ldots, T'_{b-1}, T_b)$ is shorter than $(T_a, T_{a+1}, \ldots, T_b)$, a contradiction. \qed

Observe that, having precalculated $w_\text{min}(X, Y)$ for every pair of clusters $X$ and $Y$ and $w_\text{max}(x, Y)$ and $w_\text{max}(Y, x)$ for every pair of vertex $x$ and cluster $Y$, it takes only $O(1)$ time to compute the lower bound (\ref{eq:shortest_path_lower_bound}).  A drawback of this approach is that it needs the shortest cycle $(T_1, T_2, \ldots, T_m, T_1)$ corresponding to the current solution, i.e., every time an improvement is found, one has to use \CO{} to find the tour itself (recall that we normally need only the cluster order and the weight of current solution).  These additional calls of \CO{}, however, do not take much time in practice.

In our experiments the use of the lower bound speeds up the 2-opt Global adaptation in about three times, on average.  The lower bound works better for large instances because the lower bounds for large instances have better relative precision.  Indeed, the number of edges calculated imprecisely is always fixed while the total number of edges included in the lower bound increases with the increase of the instance size.

In certain cases the lower bound (\ref{eq:shortest_path_lower_bound}) can be improved using Theorem~\ref{th:shortest_path_lower_bound_cycle}.

\begin{theorem}
\label{th:shortest_path_lower_bound_cycle}
For the shortest path from an arbitrary vertex in $\mathcal{T}_1$ to an arbitrary vertex in $\mathcal{T}_m$ in a layered network $\mathcal{T}_1 \cup \mathcal{T}_2 \cup \ldots \cup \mathcal{T}_m$ we have:
$$
w_\text{min}(\T{1}, \T{2}, \ldots, \T{m}) \ge w(T_1, T_2, \ldots, T_m, T_1) - w_\text{max}(\T{m}, \T{1}) \,,
$$
where $(T_1, T_2, \ldots, T_m, T_1)$ is the shortest cycle through all the layers of the network.
\end{theorem}
\proof
Assume that there exists a path $(T'_1, T'_2, \ldots, T'_m)$, $T'_i \in \T{i}$, such that
$$
w(T'_1, T'_2, \ldots, T'_m) < w(T_1, T_2, \ldots, T_m, T_1) - w_\text{max}(\T{m}, \T{1}) \,.
$$
Close up this path with the edge $(T'_m, T'_1)$.  Observe that the weight of the obtained cycle is
\begin{equation}
\label{eq:better_path}
w(T'_1, T'_2, \ldots, T'_m, T'_1) < w(T_1, T_2, \ldots T_m, T_1) + w(T'_m, T'_1) - w_\text{max}(\mathcal{T}_m, \mathcal{T}_1) \,.
\end{equation}
Thus, $w(T'_1, T'_2, \ldots, T'_m, T'_1) < w(T_1, T_2, \ldots, T_m, T_1)$, a contradiction. \qed

\subsubsection{Supporting Cluster}
\label{sec:supporting_cluster}

Observe that, in general, skipping some of the shortest cycles calculations (see Section~\ref{sec:lower_bound}) does not decrease the time spent to find the shortest paths.  Indeed, even if the shortest paths between some clusters \T{i} and \T{j} are not required due to the lower bound, these paths are still needed, e.g., to find the shortest paths between \T{i} and \T{j+1}.

We propose an approach that significantly reduces the number of shortest paths required for the global adaptation.  It also guarantees that the layered network $L$ constructed on every iteration will always contain the smallest cluster (recall that the \CO{} performance significantly depends on the size of the smallest cluster in $L$).  This is achieved at the cost of a larger number of layers in $L$.

\begin{figure}[ht]
\centerline{
\xymatrix@R=1.5em@C=2em@L=0.5em{
		*+{\mathcal{T}_{y+1}} \ar@3{-}[r]
	&	\txt{Shortest\\paths} \ar@3{->}[r]
	&	*+{\mathcal{T}_1} \ar@3{-}[r]
	&	\txt{Shortest\\paths} \ar@3{->}[r]
	&	*+{\mathcal{T}_x} \ar@3{-}[d]
\\
		\txt{$w(u, v)$\\$u \in \mathcal{T}_{x+1}$ and $v \in \mathcal{T}_{y+1}$} \ar@3{->}[u]
	&
	&
	&
	&	\txt{$w(u, v)$\\$u \in \mathcal{T}_x$ and $v \in \mathcal{T}_y$} \ar@3{->}[d]
\\
		*+{\mathcal{T}_{x+1}} \ar@3{-}[u]
	&
	&	\txt{Shortest\\paths} \ar@3{->}[ll]
	&
	&*+{\mathcal{T}_y} \ar@3{-}[ll]
}
}
\caption{\T{1} is a supporting cluster in the layered network $L$\@.  instead of calculating the shortest paths from every \T{y+1} to every \T{x}, i.e., for $O(m^2)$ combinations of $x$ and $y + 1$, we only need the shortest paths from every \T{y+1} to \T{1}, i.e., for $O(m)$ values of $y + 1$, and from \T{1} to every \T{x}, i.e., for $O(m)$ values of $x$.}
\label{fig:two_opt_supporting_cluster}
\end{figure}
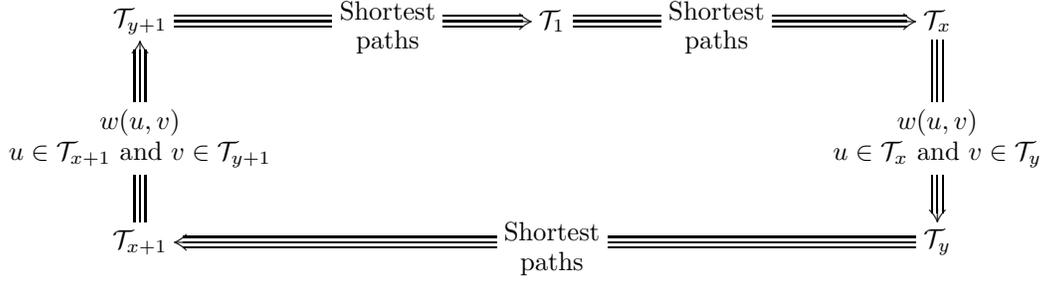

Consider the \twoopt{G}{} implementation discussed in Section~\ref{sec:two_opt_global_adaptation}.  Observe that the fragment $(\T{y+1}, \T{y+2}, \ldots, \T{x})$ always contains cluster $\T{1}$.  Let us calculate all the shortest paths in fragments $(\T{1}, \T{2}, \ldots, \T{i})$ and $(\T{i+1}, \T{i+2}, \ldots, \T{m}, \T{1})$ for every $i = 2, 3, \ldots, m - 1$.  Now, by adding \T{1} as an additional layer to the layered network $L$, we avoid calculations of the shortest paths from \T{y+1} to \T{x}, see Figure~\ref{fig:two_opt_supporting_cluster}.  We call \T{1} a \emph{supporting cluster}.

Let us find out how a supporting cluster influences the algorithm's performance.  Observe that adding an extra layer to $L$ requires $O(m n s^2)$ extra operation to calculate the shortest cycle.  However, adding an extra layer may also save some operations.  Since we are allowed to rotate the tour, let \T{1} be the smallest cluster, i.e., $|\T{1}| = \gamma$.  Then a more accurate estimation shows that the implementation of \twoopt{G}{} proposed in Algorithm~\ref{alg:two_opt_global} spends $O(m n (2s^2 + s))$ operations on all the \CO{} runs, and with a supporting cluster it would take $O(m n \gamma (3s + 1))$ operations on it.  Hence, if $\gamma / s < 2 / 3$, which is very typical, introducing the supporting cluster speeds up the algorithm.

Observe that supporting cluster can be used only if a group of fragments shares some cluster, preferably of a small size.  Next we propose an improvement of this technique that gives more flexibility and improves the time complexity of the algorithm.

\subsubsection{Multiple Supporting Clusters}
\label{sec:multiple_supporting_clusters}

Let us consider the problem of finding the shortest paths along a sequence of clusters $(\T{1}, \T{2}, \ldots, \T{m})$, i.e., finding the shortest path from every $u \in \T{i}$ to every $v \in \T{j}$ through $(\T{i+1}, \T{i+2}, \ldots, \T{j-1})$ for every $1 \le i < j \le m$.  Using the dynamic programming approach straightforwardly, one can solve the problem in $O(n s^2 m)$ time.  We propose an algorithm that, by introducing several supporting clusters, solves it in $O(n s^2 \log_2{m})$ operations such that every $(u,v)$-path contains at most one supporting cluster.

For $m = 2$, no calculations are required because the shortest $(u,v)$-path, $u \in \T{1}$ and $v \in \T{2}$, is $(u, v)$.  For $m > 2$, let us introduce a supporting cluster \T{m/2} and calculate all the shortest paths in $(\T{i}, \T{i+1}, \ldots, \T{m/2})$ and $(\T{m/2}, \T{m/2+1}, \ldots, \T{j})$ for every $i \le m / 2$ and every $j \ge m / 2$.  This takes $O(n s^2)$ operations.  Using the same technique, find the shortest paths in the subsequences $(\T{1}, \T{2}, \ldots, \T{m/2 - 1})$ and $(\T{m/2 + 1}, \T{m/2 + 2}, \ldots, \T{m})$.  Using recursion, we can solve the whole problem in $O(n \log_2{m} s^2)$ operations.  Now, in order to obtain the shortest $(u,v)$-paths, where $u \in \T{i}$, $v \in \T{j}$ and $1 \le i < j \le m$, do the following.  If either $i = m/2$ or $j = m/2$, corresponding shortest paths are already calculated.  If $i < m/2$ and $j > m / 2$, take the shortest paths from \T{i} to \T{m/2} and from \T{m/2} to \T{j} and use \T{m/2} as a supporting cluster.  If $j < m / 2$ or $i > m / 2$, refer to the corresponding subproblem.

Note that splitting the sequence of clusters into two parts is optimal.  For example, splitting it into three parts requires $O(\frac{5}{3} m n s^2)$ operations to calculate the shortest paths for these two supporting clusters, i.e., the recursive procedure takes $O(\frac{5}{3} \log_3{m} s^2)$ operations.  Note that $5 / 3 \log_3{m} > \log_2{m}$ for every $m > 1$.

\bigskip

Selecting \T{m/2} as a supporting cluster is the optimal choice when $|\T{i}| = s$ for every $i$.  In practice, it is often better to select some other cluster \T{t} such that $t \approx m/2$ if $|\T{t}| < |\T{m/2}|$.  Indeed, the size of the supporting cluster is important during both calculating the shortest paths and running \CO{}\@.  Finding the optimal $t$, however, is hard.  We use the following simple heuristic to find a good value of $t$.  We select the supporting cluster $\T{t}$ such that 
\begin{equation}
\label{eq:supporting_cluster_selection}
\displaystyle |\T{t}| = \min_{|i - m/2| \le m/6} |\T{i}| \text{ and } |t - m / 2| \text{ is minimized.}
\end{equation}

Since the positions of the supporting clusters are variable, there has to be a data structure to store them, and an algorithm is required to find the necessary supporting cluster when seeking for the shortest path between \T{i} and \T{j} for some $1 \le i < j \le m$.  For this purpose we build a binary tree of supporting cluster positions.  The root of this tree is the index $t$ of the supporting cluster selected for the sequence $(\T{1}, \T{2}, \ldots, \T{m})$.  The root has two children corresponding to the supporting clusters selected in the sequences $(\T{1}, \T{2}, \ldots, \T{t})$ and $(\T{t}, \T{t+1}, \ldots, \T{m})$, respectively, etc.  

We do not calculate all the shortest paths to and from the supporting clusters in advance but use the dynamic programming approach.  This saves significant time if some local search move is accepted.

Note that it takes $O(\log_2{m})$ operations to find the necessary supporting cluster.  However, we can usually do this search in $O(1)$ operations by reusing the result of the previous search, see Algorithm~\ref{alg:supporting_cluster_search}.
\begin{algorithm}[ht]
	\caption{Search for the supporting cluster for \twoopt{G}{}.}
	\label{alg:supporting_cluster_search}

\begin{algorithmic}
\REQUIRE Fixed position $x$ (see Algorithm~\ref{alg:two_opt_enumerate}).
\REQUIRE The supporting cluster tree defined by \var{root}, $\func{left}(i)$ and $\func{right}(i)$.

\STATE Let $k \gets 0$.
\STATE Initialize current supporting cluster position $t \gets \var{root}$.
\WHILE {$t < x$ or $t > x + 2$}
	\IF {$t > x + 2$}
		\STATE Set $k \gets k + 1$ and save $p_k \gets t$.
		\STATE $t \gets \func{left}(t)$.
	\ELSE
		\STATE $t \gets \func{right}(t)$.
	\ENDIF
\ENDWHILE

\STATE $t \gets p_k$.
\FOR {$y \gets x + 2, x + 3, \ldots, \min\{m, x + m - 2\}$}
	\IF {$k > 1$ and $y = p_{k-1}$}
		\STATE $k \gets k - 1$.
		\STATE Use the distances from \T{y} to \T{x+1}.
		\STATE Update current supporting cluster $t \gets p_k$.
	\ELSE
		\STATE Use the distances from \T{y} to \T{t} and from \T{t} to \T{x + 1} with supporting cluster \T{t}.
	\ENDIF
\ENDFOR
\end{algorithmic}
\end{algorithm}
In this algorithm, we exploit the fact that two supporting clusters can never have the same position.  Thus, the whole supporting cluster tree can be stored in an array of size $m$, and a supporting cluster can be located by its position.

With all the improvements, \twoopt{G}{} takes only $O(\gamma s n + n s^2 \log_2 m)$ operations on shortest paths calculation and $O(\gamma s m n)$ operations on running \CO{} on every iteration.  Recall that the original implementation of \twoopt{G}{} takes $O(s^2 m n)$ operations to proceed.  Hence, the time complexity of the refined \twoopt{G}{} implementation is $O(sn (s \log_2 m + \gamma m))$ which is $O\left(\frac{s}{\gamma} \right)$ times faster than $O(s^2 m n)$.

\bigskip

In the discussion above, we assumed exploration of a full neighborhood and, hence, calculated all the needed shortest paths along $(\T{1}, \T{2}, \ldots, \T{m})$.  However, in practice, we do not normally explore the whole neighborhood but rearrange the tour as soon as we find an improvement.  Hence, heavy preprocessing of a tour is usually unacceptable.  This means that we should calculate as few shortest paths as necessary for every particular candidate and when an improvement is accepted we should reuse the precalculated shortest paths as many times as possible.

We propose the following implementation.  A matrix $S_{u,v}$ is used to store the shortest distances along the given fragment, where $u$ and $v$ are the origin and the destination vertices, respectively.  There are $m - 2$ possible supporting clusters; for every possible supporting cluster \T{i} we store positions $\func{left}(i)$, $\func{right}(i)$, $\func{leftmost}(i)$ and $\func{rightmost}(i)$.  Positions $\func{left}(i)$ and $\func{right}(i)$ point to the child supporting clusters of \T{i}; $\func{leftmost}(i)$ is the position of the leftmost cluster from which the shortest distance to \T{i} are calculated and valid; $\func{rightmost}(i)$ is the position of the rightmost cluster to which the shortest distances from \T{i} are calculated and valid.

First, we initialize $\func{leftmost}(i) \gets i$, $\func{rightmost}(i) \gets i$, $\func{left}(i) \gets -1$ and $\func{right}(i) \gets -1$ for every $i$ and select the root position $\var{root}$ according to (\ref{eq:supporting_cluster_selection}).  The values $\func{left}(i)$ or $\func{right}(i)$ are then calculated on demand according to the same procedure.

In Algorithm~\ref{alg:supporting_cluster_search}, prior to using the shortest distances from cluster \T{j} to supporting cluster \T{i}, we check if $\func{leftmost}(i) \le j$.  If not, we update the shortest distances $S$ and $\func{leftmost}(i)$ accordingly.  Similarly, we use the value $\func{rightmost}(i)$ when we need the shortest distances from \T{i} to \T{j}.

When a tour fragment $(\T{x}, \T{x+1}, \ldots, \T{y})$ is modified, we update all the information for every possible supporting cluster.  In particular, in the \twoopt{G}{} implementation, if $x \le i \le y$, we reset all the corresponding information: $\func{leftmost}(i) \gets i$ and $\func{rightmost}(i) \gets i$.  Otherwise, if $\func{leftmost}(i) \le y$, we update $\func{leftmost}(i) \gets y + 1$ and if $\func{rightmost}(i) \ge x$, we update $\func{rightmost}(i) \gets x - 1$.  We also reset all the values $\func{left} = \func{right} = -1$.  Finally, we choose the root position $\var{root}$ according to the procedure above.  Note that, although we destroy the supporting cluster tree every time the tour is updated, it is likely that the new tree will reuse some of the old supporting clusters with all accumulated data.

\subsection{$k$-opt}
\label{sec:k_opt}

$k$-opt neighborhood is widely used for the TSP and some other combinatorial optimization problems, see, e.g.,~\citep{Fischetti1997,GK_MAP_LS_2010,GK_GTSP_GA_2008,Snyder2000}.  It was shown to be very efficient for the TSP \citep{Helsgaun2009}.  In general, $N_\text{$k$-opt}(T)$ contains all the solutions that can be obtained from $T$ by selecting $k$ elements in $T$ and then replacing them with $k$ new elements such that the feasibility of the solution is preserved.  In the TSP and the GTSP, $k$-opt means replacing $k$ existing edges in the solution with $k$ new edges.

The time complexity of $k$-opt increases exponentially with the growth of $k$.  In practice, only 2-opt and 3-opt are used for the TSP~\citep{Helsgaun2000,Lin1965} with rare exceptions~\citep{Helsgaun2009}.  We do not consider $k$-opt for $k > 3$.

\subsection{2-opt}
\label{sec:two_opt}

For $k = 2$ and for a fixed pair of edges $(T_x, T_{x+1})$, $(T_y, T_{y+1})$ there are only two options for every 2-opt move, i.e., to replace these edges either with $(T_x, T_y)$ and $(T_{x+1}, T_{y+1})$ or with $(T_{y+1}, T_{x+1})$ and $(T_y, T_x)$.  However, for the symmetric case both options are identical and it takes only $O(1)$ operations to evaluate a 2-opt move, see~(\ref{eq:turn_delta}).  Hence, it takes $O(m^2)$ operations to explore the whole neighborhood $N_\text{2-opt}(T)$ in the symmetric case.

We consider two algorithms to explore the 2-opt neighborhood, namely \emph{`simple'} and \emph{`advanced'}.  The `simple' one tries all feasible pairs of $x$ and $y$ with $y > x$, see Algorithm~\ref{alg:two_opt_basic}.
\begin{algorithm}[th]
\caption{Basic 2-opt algorithm, `simple' implementation (symmetric case).}
\label{alg:two_opt_basic}
\begin{algorithmic}
\REQUIRE Tour $T = (T_1, T_2, \ldots, T_m, T_1)$.
\STATE Initialize $b(T_i) \gets \text{true}$ for every $i = 1, 2, \ldots, m$.
\REPEAT
	\STATE Initialize $\var{optimal} \gets \text{true}$.
	\STATE Initialize $b'(T_i) \gets \text{false}$ for every $i = 1, 2, \ldots, m$.
	\FOR {$x \gets 1, 2, \ldots, m - 2$}
		\FOR {$y \gets x + 2, x + 3, \ldots, \min\{ m, x + m - 2 \}$}
			\IF {$b(T_x) = \text{false}$ and $b(T_y) = \text{false}$}
				\STATE Go to the next $y$.
			\ENDIF
			\STATE $\Delta \gets w(T_x, T_y) + w(T_{x+1}, T_{y+1}) - w(T_x, T_{x+1}) - w(T_y, T_{y+1})$.
			\IF {$\Delta < 0$}
				\STATE Replace edges $(T_x, T_{x+1})$ and $(T_y, T_{y+1})$ in $T$ with edges $(T_x, T_y)$ and $(T_{x+1}, T_{y+1})$.
				\STATE `Invalidate' vertices: $b'(T_i) \gets \text{true}$ for every $i = x, x + 1, \ldots, y$.
				\STATE Set $\var{optimal} \gets \text{false}$.
				\STATE Continue to the next $x$.
			\ENDIF
		\ENDFOR
	\ENDFOR
	\STATE Swap $b$ and $b'$.
\UNTIL {$\var{optimal} = \text{true}$}
\end{algorithmic}
\end{algorithm}
Note that after an improvement is applied, it is not necessary to explore the whole neighborhood again.  We use an efficient approach to avoid such repetitions.  In particular, the algorithm stores a flag $b(T_i)$ for every vertex $T_i$.  This flag shows if the edge starting from $T_i$ was changed since the last check.  Observe that a move of $\Turn(T, x, y)$ is redundant if both edges $(T_x, T_{x+1})$ and $(T_y, T_{y+1})$ stay unchanged since the last check of $\Turn(T, x, y)$.

The second, `advanced', algorithm is only suitable for symmetric problems.  It considers all the values $x \in \{ 1, 2, \ldots, m \}$ and for every $x$ it takes all feasible $y$ such that $w(T_x, T_y) < w(T_x, T_{x+1})$ or $w(T_{x+1}, T_{y+1}) < w(T_x, T_{x+1})$.  Indeed, if a pair of edges was not considered at all (neither when $x > y$ nor when $x < y$), then both $w(T_x, T_y) \ge w(T_x, T_{x+1})$ and $w(T_{x+1}, T_{y+1}) \ge w(T_y, T_{y+1})$ which cannot be an improving move.  For details see~\citep{Johnson2002}.

An efficient implementation of the `advanced' algorithm requires some precalculation.  Let $l(v)$ be a list of all vertices $v' \neq v$ ordered such that $w(v, l(v)_i) \le w(v, l(v)_j)$ for every $i < j$.  For a fixed $x$, try $T_y \gets l(T_x)_i$ for every $i = 1, 2, \ldots$ until $w(T_x, T_y) \ge w(T_x, T_{x+1})$.  Similarly, try $T_{y+1} \gets l(T_{x+1})_i$ for every $i = 1, 2, \ldots$ until $w(T_{x+1}, T_{y+1}) \ge w(T_x, T_{x+1})$.  This will exhaust all necessary values of $y$.  Note that for the GTSP, one has either to precalculate lists $l(v)$ every time before the \twoopt{}{} run or, instead, keep clusters in the lists $l(v)$ such that $w_\text{min}(v, l(v)_i) \le w_\text{min}(v, l(v)_j)$ for any $i < j$.

\bigskip

For the asymmetric problem, one standalone move $\Turn(T, x, y)$ of \twoopt{}{} requires $O(m)$ operations.  There are two options to reconnect the fragments and each of the options requires one of these fragments to be inverted.  However, it is still possible to explore the whole neighborhood $N_\text{2-opt}(T)$ in $O(m^2)$.  For this purpose the \twoopt{}{} moves should be carried out in a certain sequence, see Algorithm~\ref{alg:two_opt_asymmetric}.
\begin{algorithm}[ht]
\caption{Basic 2-opt implementation for asymmetric problem.}
\label{alg:two_opt_asymmetric}
\begin{algorithmic}
\REQUIRE Tour $T = (T_1, T_2, \ldots, T_m, T_1)$.
\FOR {$x \gets 1, 2, \ldots, m - 2$}
	\STATE Initialize $\delta \gets 0$.
	\FOR {$y \gets x + 2, x + 3, \ldots, \min\{ m, x + m - 2 \}$}
		\STATE Update $\delta \gets \delta + w(T_{y-1}, T_y) - w(T_y, T_{y - 1})$.
		\STATE $\Delta \gets w(T_x, T_y) + w(T_{x+1}, T_{y+1}) - w(T_x, T_{x+1}) - w(T_y, T_{y+1}) - \delta$.
		\IF {$\Delta < 0$}
			\STATE The tour $\Turn(T, x, y)$ is an improvement over $T$.
		\ENDIF
	\ENDFOR
\ENDFOR
\end{algorithmic}
\end{algorithm}
On every iteration, the variable $\delta$ stores the weight difference caused by inverting the fragment $(T_{x+1}, T_{x+2}, \ldots, T_y)$, i.e., 
$$
\delta = w(T_{x+1}, T_{x+2}, \ldots, T_y) - w(T_y, T_{y-1}, \ldots, T_{x+1}) \,.
$$
In order to consider the moves $\Turn(T, x, y)$ where $x > y$, invert the given tour $T_\text{inv} = (T_m, T_{m-1}, \ldots, T_1, T_m)$ and apply the procedure again.

Observe that the time complexity of Algorithm~\ref{alg:two_opt_asymmetric} is $O(m^2)$.

\bigskip

Our Local adaptation of 2-opt (\twoopt{L}{}, \twooptshort{L}{}) is based on Algorithm~\ref{alg:two_opt_basic}.  For every pair $x$ and $y$ it finds the shortest paths $(T_{x-1}, T'_x, T'_y, T_{y-1})$ and $(T_{x+2}, T'_{x+1}, T'_{y+1}, T_y)$, where $T'_i \in \Cluster(T_i)$ for $i \in \{ x, x + 1, y, y + 1 \}$.  The time complexity of \twoopt{L}{} is $O(m n s)$.

\bigskip

Our Global adaptation of 2-opt (\twoopt{G}{}, \twooptshort{G}{}) exploits all the approaches proposed in Section~\ref{sec:global}.  Some further discussion of the \twoopt{G}{} implementation performance can be found below.

Note that \twoopt{G}{} is naturally suitable for both symmetric and asymmetric problems.  However, in order to explore the whole neighborhood for an asymmetric problem, the procedure has to be applied twice: for a tour $T$ and then for an inversed tour $T_\text{inv} = (T_m, T_{m-1}, \ldots, T_1, T_m)$.

\bigskip

Table~\ref{tab:implementations_two_opt} reports the running times of two Basic and three Global adaptations of \twoopt{}{}.  \twoopt{G}{} is a fully optimized implementation that applies all the improvements discussed in Section~\ref{sec:global_refinements}.  \twoopt{G simple}{} is a simplified variation of the algorithm that constructs layered networks $L$ and applies \CO{} to them on every iteration but does not introduce any supporting clusters or lower bounds.  \twoopt{G naive}{} is a naive implementation of \twoopt{G}{} that applies \CO{} to every candidate $T' \in N_\text{2-opt}(T)$.

\begin{table}[ht] \centering
\scriptsize
\caption{Comparison of different 2-opt implementations.  The reported values are running times, in ms.}
\label{tab:implementations_two_opt}
\begin{tabular}{lrrc@{\qquad}rrr}
\toprule

&\multicolumn{2}{c}{Basic}&&\multicolumn{3}{c}{Global}\\
\cmidrule(){2-3}
\cmidrule(){5-7}
Instance&\optshort{2}{B}{}&\optshort{2}{B adv.}{}&&\optshort{2}{G}{}&\optshort{2}{G simple}{}&\optshort{2}{G naive}{}\\
\cmidrule(){1-7}

10att48&0.5&0.4&&0.3&0.5&0.1\\
12brazil58&0.0&0.2&&0.1&0.5&0.4\\
20rat99&0.0&0.1&&0.3&1.6&0.9\\
20kroe100&0.0&0.1&&0.2&1.1&0.8\\
24gr120&0.0&0.1&&0.3&3.3&1.1\\
28gr137&0.0&0.4&&0.5&4.4&3.7\\
31pr152&0.0&0.2&&0.2&3.7&3.3\\
40d198&0.1&0.5&&1.3&17.9&20.1\\
45tsp225&0.1&0.3&&1.3&13.5&20.6\\
56a280&0.1&0.5&&2.2&24.2&37.1\\
87gr431&0.1&1.1&&2.6&56.9&187.3\\
107att532&0.2&1.7&&4.2&85.4&296.5\\
131p654&0.3&2.6&&4.9&171.8&842.6\\
200dsj1000&0.9&6.8&&28.0&780.4&6942.4\\
\cmidrule(){1-7}

Average&0.2&1.1&&3.3&83.2&596.9\\
\bottomrule

\end{tabular}
\end{table}
One can see that \twoopt{B adv.}{} (the `advanced' implementation, see above) is usually inefficient for the GTSP\@.  Observe that the time required to generate lists $l(v)$ is $O(m^2 \log{m})$ while it takes only $O(m^2)$ operations to explore the whole neighborhood $N_\text{2-opt}(T)$ with the `simple' algorithm.  To speed up the precalculation part, we tried to include in $l(v)$ only the closest to $v$ vertices but with no success.  We assume that \twoopt{B adv.}{} may be useful as a part of a powerful metaheuristic that needs to run \twoopt{}{} many times for one instance.

As regards the Global implementations, it follows from Table~\ref{tab:implementations_two_opt} that, on average, \twoopt{G}{} is more than 10 times faster than \twoopt{G simple}{} and more than 100 times faster than \twoopt{G naive}{}.  Note that the speed-up highly depend on $m$ and is better visible for large instances.  This is because \twoopt{G simple}{} is $\Theta(m \gamma / s)$ times faster than \twoopt{G naive}{} and also because the lower bound in \twoopt{G}{} is very efficient when $m$ si large, see Section~\ref{sec:lower_bound}.

\bigskip

Different adaptations of \twoopt{}{} are compared in Table~\ref{tab:two_opt}.
\begin{table}[ht] \centering
\scriptsize
\caption{\twoopt{}{} adaptations comparison.}
\label{tab:two_opt}
\begin{tabular}{lcrrrrrcrrrrr}
\toprule

&&\multicolumn{5}{c}{Solution error, \%}&&\multicolumn{5}{c}{Running time, ms}\\
\cmidrule(){3-7}
\cmidrule(){9-13}
Instance&&\optshort{2}{B}{}&\optshort{2}{B}{co}&\optshort{2}{L}{}&\optshort{2}{L}{co}&\optshort{2}{G}{}&&\optshort{2}{B}{}&\optshort{2}{B}{co}&\optshort{2}{L}{}&\optshort{2}{L}{co}&\optshort{2}{G}{}\\
\cmidrule(){1-13}

10att48&&8.5&6.3&2.3&2.3&2.3&&0.52&0.22&0.21&0.21&0.28\\
12brazil58&&14.0&2.1&4.2&1.5&1.1&&0.01&0.01&0.01&0.02&0.08\\
20rat99&&22.1&17.1&16.5&13.7&0.8&&0.01&0.05&0.03&0.04&0.33\\
20kroe100&&15.2&1.3&5.4&2.7&0.0&&0.01&0.03&0.03&0.03&0.18\\
24gr120&&30.2&16.8&9.1&10.3&15.2&&0.01&0.03&0.06&0.10&0.27\\
28gr137&&9.6&1.9&3.6&2.7&1.9&&0.02&0.05&0.05&0.06&0.46\\
31pr152&&9.8&4.1&6.6&2.4&1.3&&0.02&0.03&0.04&0.06&0.21\\
40d198&&7.3&8.7&3.8&5.0&1.5&&0.05&0.14&0.14&0.28&1.34\\
45tsp225&&20.8&14.0&12.0&9.4&6.8&&0.05&0.15&0.13&0.23&1.26\\
56a280&&26.9&13.3&18.9&10.8&14.6&&0.06&0.15&0.19&0.30&2.17\\
87gr431&&10.3&4.8&8.7&6.9&4.2&&0.14&0.48&0.37&0.52&2.63\\
107att532&&16.8&9.2&16.1&14.2&7.9&&0.22&0.69&0.58&1.02&4.21\\
131p654&&4.1&6.9&9.0&7.7&4.0&&0.33&1.42&0.74&1.48&4.88\\
200dsj1000&&23.3&12.9&17.9&16.1&12.9&&0.91&3.27&3.11&5.28&28.04\\
\cmidrule(){1-13}

Average&&15.6&8.5&9.6&7.5&5.3&&0.17&0.48&0.41&0.69&3.31\\
\bottomrule
\end{tabular}
\end{table}
We measure \emph{solution error} as $e(T) = \frac{w(T) - w(T_\text{optimal})}{w(T_\text{optimal})} \cdot 100\%$, where $T_\text{optimal}$ is the optimal solution.

The Basic adaptation \twoopt{B}{} is the fastest but also the weakest one.  It takes only 1~ms to proceed even for the largest instances, however, it is not able to change vertex selection which makes its solution quality noncompetitive.  The \twoopt{B}{co} and \twoopt{L}{} adaptations, thus, are significantly better with respect to solution quality.  The most powerful adaptation \twoopt{G}{} is only about five times slower than the next powerful one \twoopt{L}{co} although the neighborhood of \twoopt{G}{} is significantly larger than the one of \twoopt{L}{co}.  This shows again the efficiency of the refinements proposed in Section~\ref{sec:global_refinements}.

\subsection{3-opt}
\label{sec:three_opt}

After removing edges $(T_x, T_{x+1})$, $(T_y, T_{y+1})$ and $(T_z, T_{z+1})$ from a tour $T$, depending on the symmetry of the problem, we get four or eight options to reconnect the tour fragments to obtain a feasible tour $T'$ such that $(T_x, T_{x+1}),\, (T_y, T_{y+1}),\, (T_z, T_{z+1}) \notin T'$\@.  However, we limit ourselves to only one of these options, which does not turn any of the tour fragments.  Note that all the other options can be replaced with sequences of two non-independent 2-opt moves~\citep{Rego2006} such as $\Turn(\Turn(T, x, y), x, z)$ or $\Turn(\Turn(T, x, y), y, z)$.

We implemented all the adaptations (see Section~\ref{sec:ls_adaptation}) of the 3-opt neighborhood and found out that the obtained algorithms are rather slow than powerful.  However, it is worth noting that the Global adaptation for 3-opt can be implemented quite efficiently.  Indeed, it takes $O(n s^2 \log_2 m)$ time to find the shortest paths from every vertex $u$ to every vertex $v \notin \Cluster(u)$ along the tour, see Sections~\ref{sec:supporting_cluster} and~\ref{sec:multiple_supporting_clusters}.  Then, it takes only $O(\gamma s m^2 n)$ time to perform cluster optimization for all the triples $x$, $y$, $z$.  Hence, the whole algorithm's time complexity is $O(s n (s \log_2 m + \gamma m^2))$ which is at most $O(m)$ times slower than \twoopt{G}{}.  In addition, one can apply the lower bound for the shortest cycle (see Theorem~\ref{th:shortest_path_lower_bound}) which significantly sped-up the algorithm in our experiments.

\subsection{Insertion}

The \emph{Insertion} TSP neighborhood includes all the solutions which can be obtained from the given one by removing a vertex and inserting it into some other position.  Observe that $N_\text{ins}(T) \subset N_\text{3-opt}(T)$ (consider 3-opt where one of the fragments consists of exactly one vertex).  The size of the Insertion neighborhood is $|N_\text{ins}(T)| = m(m - 2)$.

We implemented all the adaptations (see Section~\ref{sec:ls_adaptation}) for Insertion (\insertion{}{}).  As a quick improvement (\func{QuickImprove}) for the local adaptations \insertion{L}{} and \insertion{L}{co}, we optimize the vertices within inserted cluster and two clusters around its old position.  For a lower bound in the Global adaptation (\insertion{G}{}) we use the results of Theorem~\ref{th:shortest_path_lower_bound_cycle}.

Some of these adaptations have already been used in the literature.  For example, \insertion{L}{} was used by \citet{Snyder2000} (it is called \emph{Swap} there) and by \citet{Renaud1998} (\emph{G-opt} heuristic).  The \emph{Move} heuristic by \citet{Bontoux2009} is \insertion{G}{}.  However, in~\citep{Bontoux2009} the neighborhood is explored with a heuristic algorithm which does not guarantee that it finds a local minimum.

\bigskip

In Table~\ref{tab:insertion}, we provide experimental results for all the adapations of \insertion{}{}.
\begin{table}[ht] \centering
\scriptsize
\caption{\insertion{}{} adaptations comparison.}
\label{tab:insertion}
\begin{tabular}{lcrrrrrcrrrrr}
\toprule

&&\multicolumn{5}{c}{Solution error, \%}&&\multicolumn{5}{c}{Running time, ms}\\
\cmidrule(){3-7}
\cmidrule(){9-13}
Instance&&\insertion{B}{}&\insertion{B}{co}&\insertion{L}{}&\insertion{L}{co}&\insertion{G}{}&&\insertion{B}{}&\insertion{B}{co}&\insertion{L}{}&\insertion{L}{co}&\insertion{G}{}\\
\cmidrule(){1-13}

10att48&&4.7&2.4&0.9&0.9&0.0&&0.50&0.21&0.21&0.21&0.31\\
12brazil58&&14.0&2.1&14.5&0.1&0.0&&0.01&0.01&0.01&0.01&0.13\\
20rat99&&32.0&16.5&13.1&11.1&0.0&&0.01&0.05&0.04&0.05&1.07\\
20kroe100&&18.5&7.7&14.0&9.3&6.6&&0.01&0.03&0.03&0.06&0.72\\
24gr120&&35.1&20.7&6.4&9.1&2.0&&0.02&0.03&0.05&0.09&0.81\\
28gr137&&9.6&8.2&12.1&2.3&0.0&&0.02&0.03&0.04&0.09&1.25\\
31pr152&&12.6&8.5&8.0&7.1&5.9&&0.03&0.05&0.08&0.08&0.49\\
40d198&&25.6&21.3&14.2&20.6&15.8&&0.04&0.12&0.16&0.27&2.78\\
45tsp225&&36.2&33.1&22.5&21.5&15.2&&0.05&0.10&0.14&0.26&3.16\\
56a280&&31.9&22.3&26.8&23.4&20.9&&0.07&0.12&0.18&0.16&4.83\\
87gr431&&11.0&7.8&10.1&8.5&6.7&&0.16&0.47&0.42&0.56&6.71\\
107att532&&22.4&16.7&15.5&15.2&11.6&&0.29&0.59&0.63&1.16&15.43\\
131p654&&23.0&22.7&23.7&22.5&19.0&&0.48&1.82&0.96&2.31&20.38\\
200dsj1000&&40.7&31.1&29.1&26.6&27.4&&1.09&3.07&3.08&6.90&71.71\\
\cmidrule(){1-13}

Average&&22.7&15.8&15.1&12.7&9.4&&0.20&0.48&0.43&0.87&9.27\\
\bottomrule
\end{tabular}
\end{table}
One can see the same tendency here as in \twoopt{}{} adaptations.  Despite their quite different implementations, \insertion{B}{co} and \insertion{L}{} have very similar performance.  The Basic adaptation is extremely fast but of poor solution quality.   \insertion{L}{co} produces slightly better solutions in roughly twice larger times.  \insertion{G}{} is significantly slower than \insertion{L}{co} but its solution quality is noticeably better, especially for the small instances.

\subsection{Swap}

\begin{figure}[ht]
\centering
\subfloat[Original tour $T$.]
{
\xymatrix@R=4em@C=2.5em@L=0.5em{
		*+{T_{x-1}} \ar@{->}[r]
	&	*+{T_x} \ar@{->}[r]
	&	*+{T_{x+1}} \ar@(r,r)@{-->}[d]
\\
		*+{T_{y+1}} \ar@(l,l)@{-->}[u] 
	&	*+{T_y} \ar@{->}[l]
	&	*+{T_{y-1}} \ar@{->}[l]
}
}
\qquad\qquad\qquad
\subfloat[The tour $T$ after swapping $T_x$ and $T_y$.]
{
\xymatrix@R=4em@C=2.5em@L=0.5em{
		*+{T_{x-1}} \ar@(r,l)@{->}[rd]
	&	*+{T_x} \ar@(l,r)@{->}[ld]|\hole
	&	*+{T_{x+1}} \ar@(r,r)@{-->}[d]
\\
		*+{T_{y+1}} \ar@(l,l)@{-->}[u] 
	&	*+{T_y} \ar@(r,l)@{->}[ru]
	&	*+{T_{y-1}} \ar@(l,r)@{->}[lu]|\hole
}
}
\caption{A Swap move.}
\label{fig:swap_move}
\end{figure}
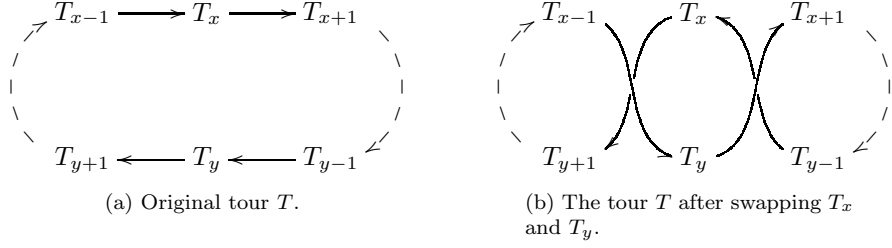

The \emph{Swap} TSP neighborhood $N_\text{swap}(T)$ contains all the solutions obtained from tour $T$ by swapping two vertices in it, see Figure~\ref{fig:swap_move}.  Observe that $|N_\text{swap}(T)| = m (m - 1)$.

An important message is that \Swap{} does not work well for near-optimal solutions.  Indeed, a \Swap{} move can be replaced with a sequence of two \insertion{}{} or \twoopt{}{} moves.  Moreover, the following theorem proves that a \twoopt{}{} local minimum is also a \Swap{} local minimum for a symmetric TSP.

\begin{theorem}
\label{th:swap}
Let $T$ be a local minimum in $N_\text{2-opt}(T)$.  Then $T$ is also a local minimum in $N_\text{swap}(T)$ if the problem is symmetric.
\end{theorem}
\proof
Assume that the tour $T$ is a local minimum in $N_\text{2-opt}(T)$ but it is not a local minimum in $N_\text{swap}(T)$.  Then, there exist some $x$ and $y$ such that $w(T') < w(T)$, where $T'$ is a tour obtained $T$ by swapping $T_x$ and $T_y$ (see Figure~\ref{fig:swap_move}):
\begin{equation}
\label{eq:swap_improves}
w(T_{x-1}, T_y, T_{x+1}) + w(T_{y-1}, T_x, T_{y+1}) < w(T_{x-1}, T_x, T_{x+1}) + w(T_{y-1}, T_y, T_{y+1}) \,.
\end{equation}

Let us consider two tours: $A = \Turn(T, x - 1, y)$ and $B = \Turn(T, x, y - 1)$.  (Without loss of generality, one may assume that $x < y$.)  According to (\ref{eq:turn_delta}),
$$
w(A) = w(T) + w(T_{x-1}, T_y) + w(T_x, T_{y+1}) - w(T_{x-1}, T_x) - w(T_y, T_{y+1}) \text{ and}
$$
$$
w(B) = w(T) + w(T_x, T_{y-1}) + w(T_{x+1}, T_y) - w(T_x, T_{x+1}) - w(T_{y-1}, T_y) \,.
$$

If $T$ is a local minimum in $N_\text{2-opt}(T)$, then both $w(A) - w(T)$ and $w(B) - w(T)$ are non-negative and their sum is also non-negative.  Since we consider a symmetric problem,
\begin{multline*}
[w(A) - w(T)] + [w(B) - w(T)] = \big[ w(T_{x-1}, T_y) + w(T_x, T_{y+1}) - w(T_{x-1}, T_x) - w(T_y, T_{y+1}) \big] \\
+ \big[ w(T_x, T_{y-1}) + w(T_{x+1}, T_y) - w(T_x, T_{x+1}) - w(T_{y-1}, T_y) \big] \\
 = \big[ w(T_{x-1}, T_y, T_{x+1}) + w(T_{y-1}, T_x, T_{y+1}) \big] - \big[ w(T_{x-1}, T_x, T_{x+1}) + w(T_{y-1}, T_y, T_{y+1}) \big] \,.
\end{multline*}
However, according to (\ref{eq:swap_improves}), this expression is negative and, hence, our assumption is wrong and the tour $T$ is a local minimum in $N_\text{swap}(T)$.
\qed

Note that this result was also observed empirically by \citet{GK_GTSP_GA_2008}.

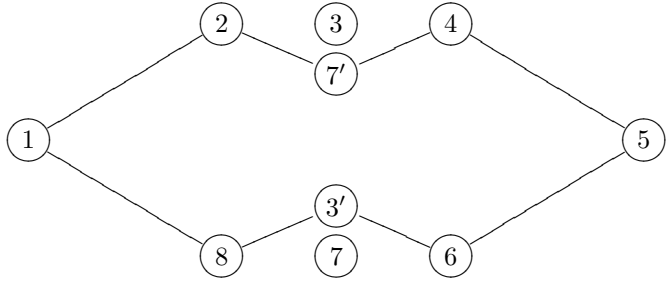
\begin{figure}[ht]
\centerline{
\xymatrix@-1pc@R=0pt@C=25pt{
	&	
	& *++[o][F-]{2} \ar@{-}[rd]
	& *++[o][F-]{3} 
	& *++[o][F-]{4} \ar@{-}[rrddd] 
\\
	&
	&
	&	*++[o][F-]{7'} \ar@{-}[ru]
\\
\\
	  *++[o][F-]{1} \ar@{-}[rruuu] \ar@{-}[rrddd]
	&
	&
	&
	&
	&
	&	*++[o][F-]{5}
\\
\\
	&
	&
	&	*++[o][F-]{3'} \ar@{-}[rd]
\\
	&	
	& *++[o][F-]{8} \ar@{-}[ru]
	& *++[o][F-]{7} 
	& *++[o][F-]{6} \ar@{-}[rruuu]
}
}
\caption{An example of a local minimum in $N_\text{2-opt G}(T)$ which is not a local minimum in $N_\text{swap L}(T)$.}
\label{fig:swap_example}
\end{figure}

Until now, we considered only the TSP Swap neighborhood.  Obviously, this result can be extended to the Basic adaptation but it is unclear if it holds for the Local and Global adaptations.
\begin{theorem}
The result of Theorem~\ref{th:swap} does not hold for the Local or Global adaptations of Swap, i.e., a local minimum in $N_\text{2-opt G}(T)$ is not necessarily a local minimum in $N_\text{swap L}(L)$ even if the problem is planar with Euclidean distances.
\end{theorem}
\proof
We will show an example of a GTSP tour $T$ which is a local minimum in $N_\text{2-opt G}(T)$ but not a local minimum in $N_\text{swap L}(T)$.  Consider an example on Figure~\ref{fig:swap_example}.
It is a planar GTSP with Euclidean distances and 8 clusters: $\{ 1 \}$, $\{ 2 \}$, $\{ 3, 3' \}$, $\{ 4 \}$, $\{ 5 \}$, $\{ 6 \}$, $\{ 7, 7' \}$ and $\{ 8 \}$.  The original tour $T$ is $T = (1, 2, 7', 4, 5, 6, 3', 8, 1)$.  Observe that swapping $3'$ and $7'$ together with optimizing the swapped vertices (i.e., replacing $3'$ and $7'$ with $3$ and $7$, respectively) produces the optimal tour $(1, 2, 3, 4, 5, 6, 7, 8, 1)$.  At the same time, no adaptation of \twoopt{}{} is able to improve $T$ because whatever is the vertex selection, any \twoopt{}{} move will yield a tour with two intersecting (and, hence, long) edges.
\qed

\section{Fragment Optimization}
\label{sec:fragment_optimization}

All the adaptations of the TSP local searches discussed in Section~\ref{sec:tsp_neighborhoods} are intended to improve the whole tour structure.  In this section we discuss local improvements.  In other words, the neighborhoods below consist of the tours that can be obtained from the original one by altering only a small fragment of it.

One can think of many kinds of fragment optimization, but we focus only on the most powerful option, i.e., a neighborhood containing all possible rearrangements in a fragment of some fixed length $k$.  Consider a tour $T = (T_1, T_2, \ldots, T_m, T_1)$.  Let $a = T_m$, $b = T_{k + 1}$, $\Omega_i = \Cluster(T_{i})$ for $i = 1, 2, \ldots, k$ and $\Omega = \{ \Omega_1, \Omega_2, \ldots, \Omega_k \}$.  Let $\func{FO}(a, b, \Omega)$ be the set of all paths from the vertex $a$ to the vertex $b$ through all the clusters in $\Omega$ being taken in an arbitrary order.  Note that $|\func{FO}(a, b, \Omega)| \in O(k! s^k)$.\footnote{The two algorithms below show that the Fragment Optimization problem is fixed-parameter tractable with respect to the parameter $k$. From the theoretical point of view, the second algorithm is more efficient than the first one, but experiments described later on show that for very small values of $k$ the first algorithm is actually faster. For more information on fixed-parameter tractability see, e.g., \citep{Downey1999,Niedermeier2006}.}

Using the routine for finding the shortest paths in a layered network (see Section~\ref{sec:cluster_optimization}), one can find the best path among $\func{FO}(a, b, \Omega)$ in $O(k! \cdot (k-1) s^2)$ operations.  In this paper, we propose two algorithms $\mathcal{F}_1$ and $\mathcal{F}_2$ that find the best path in $\func{FO}(a, b, \Omega)$ in $O(s^2 k!)$ and $O(s^2 k^2 2^k)$ time, respectively.

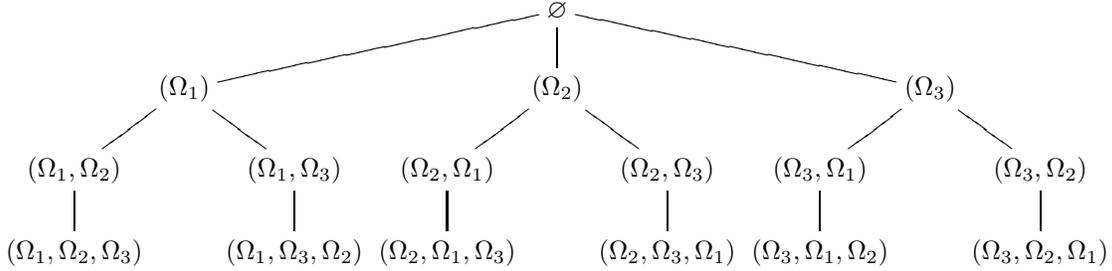
\begin{figure}
\centerline
{
\xymatrix@R=1.5em@C=0em@L=0.5em{
	&
	&
	&
	&	*+{\varnothing} \ar@{-}[llld] \ar@{-}[d] \ar@{-}[rrrd]
\\
	&	*+{(\Omega_1)} \ar@{-}[ld] \ar@{-}[rd]
	&
	&
	&	*+{(\Omega_2)} \ar@{-}[ld] \ar@{-}[rd]
	&
	&
	&	*+{(\Omega_3)} \ar@{-}[ld] \ar@{-}[rd]
\\
		*+{(\Omega_1, \Omega_2)} \ar@{-}[d]
	&
	&	*+{(\Omega_1, \Omega_3)} \ar@{-}[d]
	&	*+{(\Omega_2, \Omega_1)} \ar@{-}[d]
	&
	&	*+{(\Omega_2, \Omega_3)} \ar@{-}[d]
	&	*+{(\Omega_3, \Omega_1)} \ar@{-}[d]
	&
	&	*+{(\Omega_3, \Omega_2)} \ar@{-}[d]
\\
		*+{(\Omega_1, \Omega_2, \Omega_3)}
	&
	&	*+{(\Omega_1, \Omega_3, \Omega_2)}
	&	*+{(\Omega_2, \Omega_1, \Omega_3)}
	&
	&	*+{(\Omega_2, \Omega_3, \Omega_1)}
	&	*+{(\Omega_3, \Omega_1, \Omega_2)}
	&
	&	*+{(\Omega_3, \Omega_2, \Omega_1)}
}
}
\caption{An example of a search tree of the $\mathcal{F}_1$ algorithm for $k = 3$}
\label{fig:fragment_optimization_tree}
\end{figure}

The $\mathcal{F}_1$ algorithm is a branch and bound algorithm.  Let $S(v)$ be a sequence of distinct clusters selected from $\Omega$ assigned to search tree node $v$.  Then $S(p) = (S(v)_1, S(v)_2, \ldots, S(v)_{|S(v)| - 1})$ if $p$ is the parent node of $v$.  Set $S(\func{root}) = \varnothing$.  For an example, see Figure~\ref{fig:fragment_optimization_tree}.

Let $C = S(v)$ and $\{ x_1, x_2, \ldots, x_c \} = C_{|C|}$ be the last cluster in $C$.  For $i = 1, 2, \ldots, c$, let $l(v)_i$ be the weight of the shortest path from $a$ to $x_i$ through $C_1$, $C_2$, \ldots, $C_{|C| - 1}$.  For $i = 1, 2, \ldots, c$, let $l(v)_i = w(a, x_i)$ if $|C| = 1$.  Otherwise, if $p$ is the parent node of $v$, $P = S(p)$, $\{ y_1, y_2, \ldots, y_{c'} \} = P_{|P|}$ and we know $l(p)_j$ for every $j = 1, 2, \ldots, c'$, let $l(v)_i = \min_{j = 1, 2, \ldots, c'} l'(p)_j + w(y_j, x_i)$ for every $i = 1, 2, \ldots, c$.  If $|C| = |\Omega|$, i.e., $v$ is a tree leaf, we also calculate the shortest path from $a$ to $b$ as follows: $\min_{i = 1, 2, \ldots, c} l(v)_i + w(x_i, b)$.

The search tree contains $\sum_{i=0}^k \frac{k!}{(k - i)!} < k! \cdot e$ nodes.  It takes $O(s^2)$ to calculate the weights $l(v)$ for a node $v$.  Hence, the time complexity of $\mathcal{F}_1$ is $O(s^2 k!)$.

We can improve the performance of $\mathcal{F}_1$ by calculating the lower bound at every node $v$.  Let $l_\text{min} = \min_{i = 1, 2, \ldots, c} l(v)_i$.  Let $\lambda = \min_{X, Y \in R,\, X \neq Y} w_\text{min}(X, Y)$, where $R = \Omega \setminus P \cup \{ \Cluster(b) \}$.  Then, if $l_\text{min} + \lambda (|\Omega| - |P|) \ge w_\text{best}$, where $w_\text{best}$ is the weight of the shortest $(a,b)$-path found so far, the node $v$ and its branch are discarded.

\bigskip

The second algorithm $\mathcal{F}_2$ is preferable for large values of $k$.  It is a dynamic programming algorithm that combines the idea of the Held and Karp's TSP algorithm~\citep{Papadimitriou1982} with finding the shortest path in a layered network.  Let $\Delta \subset \Omega$ be a subset of the given clusters.  We wish to find the shortest path $p_x^\Delta$ from $a$ to every $x \notin \bigcup_{\Omega_i \in \Delta} \Omega_i$ via all the clusters $\Delta$ taken in an arbitrary order.  Observe that $p_x^\varnothing = w(a, x)$.  Assume that, for every $Y \in \Delta$, we know the shortest paths $p_y^{\Delta \setminus \{ Y \}}$ from $a$ to every $y \in Y$ through clusters $\Delta \setminus \{ Y \}$.  Then 
$$
p_x^\Delta = \min_{Y \in \Delta} \min_{y \in Y} \left\{ p_y^{\Delta \setminus \{ Y \}} + w(y, x) \right\} \,.
$$
Hence, having the required information, one can find the shortest path from $a$ to $x$ via clusters $\Delta$ taken in an arbitrary order in $O(|\Delta| s)$ time.  Observe that for $\Delta = \Omega$ and $x = b$ the algorithm finds the shortest path from $a$ to $b$ via all the clusters in the fragment.

There are $\binom{k}{|\Delta|}$ possible subsets of clusters $\Delta$ of a given size and for every subset there are $O((k - |\Delta|) s)$ vertices $x$.  It takes $O(|\Delta| s)$ operations to find each of these shortest paths.  Thus, the whole procedure takes 
$$
O\left(\sum_{|\Delta| = 1}^k \binom{k}{|\Delta|} \cdot (k - |\Delta|) s \cdot |\Delta| s\right) = O(s^2 k^2 2^k) \text{ operations.}
$$

Hence, for small values of $k$, the first algorithm $\mathcal{F}_1$ is preferable while the second algorithm $\mathcal{F}_2$ is faster for large fragments.

\bigskip

The $N_\text{$k$-FO}(T)$ neighborhood includes all the tours that can be obtained from $T$ by reordering any $k$ consequent vertices and, maybe, replacing these vertices with some other vertices from the corresponding clusters.  Let $\Phi_i^k(T)$ be a set of all tours that can be obtained from $T$ by rearranging and `reselecting' vertices $T_{i+1}$, $T_{i+2}$, \ldots, $T_{i+k}$ within the corresponding cluster.  Then $N_\text{$k$-FO}(T) = \bigcup_{i=1}^m \Phi_i^k(T)$, and to explore this neighborhood we can run either the $\mathcal{F}_1$ or $\mathcal{F}_2$ algorithm $m$ times.  Observe that $|\Phi_i^k(T) \cap \Phi_j^k(T)| \gg 1$ for some $i$ and $j$ and, hence, our algorithm explores some of the candidates in $N_\text{$k$-FO}(T)$ more than once.  It is a natural question if avoiding multiple evaluations of these candidates can save any noticeable time.

Let $A_i^k(T) = \{ T' \in \Phi_i^k(T) :\ T'_{i+1} \neq T_{i+1} \}$.  We assume that $k \le m / 2$.  Then observe that $A_i^k(T) \cap A_j^k(T) = \varnothing$ for any $i \neq j$.  Indeed, if some $T' \in A_i^k(T) \cap A_j^k(T)$ then $T'_{i+1} \neq T_{i+1}$ and $T'_{j+1} \neq T_{j+1}$.  Since $T' \in A_j^k(T)$ and the vertex $T'_{i+1}$ is modified, we get $j < i + 1 \le j + k$.  At the same time, since $T' \in A_i^k(T)$ and the vertex $T'_{j+1}$ is modified, $i < j + 1 \le i + k$.  This is only possible if $i = j$.

Observe that 
$$
\bigcup_{i=1}^m \Phi_i^k(T) \subseteq \{ T \} \cup \bigcup_{i=1}^m A_i^k(T) \,.
$$
Indeed, if $T' \in \Phi_i^k(T)$ for some $i$, then either $T' = T$ or there exists $i < j \le i + k$ such that $T'_j \neq T_j$ and $T'_p = T_p$ for every $p = i + 1, i + 2, \ldots, j - 1$.  In the latter case $T' \in A_j^k(T)$.  At the same time, 
$$
\{ T \} \cup \bigcup_{i=1}^m A_i^k(T) \subseteq \bigcup_{i=1}^m \Phi_i^k(T)
$$
since $A_i^k(T) \subset \Phi_i^k(T)$ and $T \in \Phi_i^k(T)$ for any $i$.  Hence, 
$$
\{ T \} \cup \bigcup_{i=1}^m A_i^k(T) = \bigcup_{i=1}^m \Phi_i^k(T) = N_\text{$k$-FO}(T) \,.
$$  
Recall that $A_i^k(T) \cap A_j^k(T) = \varnothing$ and observe that $|A_i^k(T)| = O\left((ks - 1) s^{k-1} (k - 1)! \right)$.  Hence, $|N_\text{$k$-FO}(T)| = O\left(m (ks - 1) s^{k-1} (k - 1)! \right)$.  

Compare it to $O(m s^k k!)$, which is the number of candidates considered by $m$ runs of either $\mathcal{F}_1$ or $\mathcal{F}_2$.  The difference is only in $\Theta(\frac{ks}{ks - 1})$ times.  We conclude that this relatively small overhead is not worth further complication of the algorithm.

Let \FO{k} be a local search with the $N_\text{$k$-FO}(T)$ neighborhood.  Then, depending on the implementation, its time complexity is either $O(m k! s^2)$ or $O(m k^2 2^k s^2)$.  

Although we know that $\mathcal{F}_1$ is more efficient for small values of $k$ and vice versa, empirical evaluation is required in order to find which algorithm is more efficient for particular values of $k$.  We compare these implementations in Table~\ref{tab:fo_implementations}.
\begin{table}[ht] \centering
\scriptsize
\caption{\FO{} implementations comparison.  The reported values are running times, in ms.}
\label{tab:fo_implementations}
\begin{tabular}{lrrrrrcrrrrr}
\toprule

&\multicolumn{5}{c}{Algorithm 1}&&\multicolumn{5}{c}{Algorithm 2}\\
\cmidrule(){2-6}
\cmidrule(){8-12}
Instance & $k = 3$ & $k = 4$ & $k = 5$ & $k = 6$ & $k = 7$ && $k = 3$ & $k = 4$ & $k = 5$ & $k = 6$ & $k = 7$ \\
\cmidrule(){1-12}

10att48&0.6&0.3&0.6&1.9&7.0&&0.3&0.4&0.6&1.3&2.7\\
12brazil58&0.0&0.2&0.6&2.2&8.8&&0.1&0.2&0.6&1.6&3.9\\
20rat99&0.1&0.3&1.2&3.5&13.5&&0.1&0.4&1.3&2.9&7.4\\
20kroe100&0.1&0.2&0.8&4.0&17.4&&0.2&0.4&1.0&3.4&8.9\\
24gr120&0.1&0.4&1.3&4.8&20.4&&0.2&0.5&1.3&3.5&9.1\\
28gr137&0.1&0.3&1.4&5.4&22.5&&0.1&0.5&1.6&4.7&13.5\\
31pr152&0.2&0.4&1.2&3.7&14.4&&0.2&0.6&1.6&3.7&9.7\\
40d198&0.2&0.5&1.9&6.8&43.4&&0.3&0.8&2.3&6.4&23.8\\
45tsp225&0.2&0.7&2.5&9.1&55.9&&0.3&1.0&2.7&7.5&27.1\\
56a280&0.2&0.7&2.2&8.0&34.5&&0.3&1.0&2.9&8.1&24.8\\
87gr431&0.5&1.7&6.4&25.3&96.2&&0.9&2.5&7.6&20.9&55.2\\
107att532&0.6&1.7&5.5&18.5&80.1&&0.8&2.4&7.1&20.0&62.0\\
131p654&0.9&2.4&8.4&31.0&125.0&&1.4&3.7&10.9&33.4&85.2\\
200dsj1000&1.4&3.5&10.1&35.3&125.2&&1.9&4.9&13.1&35.7&95.6\\
\cmidrule(){1-12}

Average&0.4&0.9&3.1&11.4&47.5&&0.5&1.4&3.9&10.9&30.6\\
\bottomrule
\end{tabular}
\end{table}
From there we see that the first implementation is faster for $k < 6$ while for $k > 6$ the second implementation is preferable, and this result holds for all the instances.  For $k = 6$ both implementations perform similarly but the second one is slightly faster on average.  Hence, in what follows, we use $\mathcal{F}_1$ when $k \le 5$ and $\mathcal{F}_2$ otherwise.

In Table~\ref{tab:fo} we provide results of experimental evaluation for the \FO{} algorithm.
\begin{table}[ht] \centering
\scriptsize
\caption{\FO{} performance for different values of $k$.}
\label{tab:fo}
\begin{tabular}{lcrrrrrcrrrrr}
\toprule

&&\multicolumn{5}{c}{Solution error, \%}&&\multicolumn{5}{c}{Running time, ms}\\
\cmidrule(){3-7}
\cmidrule(){9-13}
Instance&&\FO{2}&\FO{4}&\FO{6}&\FO{8}&\FO{10}&&\FO{2}&\FO{4}&\FO{6}&\FO{8}&\FO{10}\\
\cmidrule(){1-13}

10att48&&8.2&0.0&0.0&0.0&---&&0.5&0.3&1.3&6.4&---\\
12brazil58&&2.1&0.0&0.0&0.0&0.0&&0.0&0.2&1.6&9.1&38.8\\
20rat99&&18.5&17.9&6.8&0.8&0.0&&0.0&0.3&3.0&43.8&171.6\\
20kroe100&&24.3&24.3&23.4&23.4&0.0&&0.0&0.2&3.3&23.3&140.4\\
24gr120&&34.6&11.9&12.4&0.0&0.0&&0.0&0.5&3.6&44.9&171.6\\
28gr137&&15.0&12.8&2.4&2.1&6.3&&0.0&0.3&4.7&34.8&171.6\\
31pr152&&11.0&7.2&7.2&0.7&0.7&&0.0&0.4&3.7&34.8&202.8\\
40d198&&29.6&24.7&23.9&15.9&4.5&&0.1&0.6&6.3&67.2&468.0\\
45tsp225&&43.8&39.5&31.5&24.4&12.3&&0.1&0.7&7.5&51.8&436.8\\
56a280&&25.4&25.2&21.6&18.1&18.1&&0.1&0.7&8.1&73.1&421.2\\
87gr431&&11.1&7.6&6.6&6.0&5.9&&0.2&1.7&21.2&140.5&826.9\\
107att532&&24.0&22.4&21.7&20.6&13.5&&0.3&1.7&19.8&140.5&1123.4\\
131p654&&33.4&31.3&29.4&27.3&26.6&&0.4&2.4&33.2&218.6&1419.8\\
200dsj1000&&43.7&39.1&37.4&35.9&34.3&&0.6&3.5&35.6&250.0&1669.6\\
\cmidrule(){1-13}

Average&&23.2&18.9&16.0&12.5&9.4&&0.2&1.0&10.9&81.4&558.7\\
\bottomrule
\end{tabular}
\end{table}
It is predictable that the heuristic yields very good solutions for small instances, i.e., when $k$ is close to $m$.  On average, however, solution quality of \FO{}{} is relatively low.  We conclude that \FO{} neighborhood is more interesting in combination with some other neighborhoods than as a stand-alone heuristic.  Combining several neighborhoods, however, is a subject of a separate research.

\section{Data Structures}
\label{sec:data_structures}

Apart from the theoretical properties of an algorithm, implementation details may also have great influence on its performance.  In this section, we discuss what data structures are the most efficient and convenient for a GTSP heuristic.

\subsection{Tour Representation}

It is a non-trivial question how one should store a GTSP solution.  The most common approach is to store a sequence of vertices in the visiting order.  It was used by \citet{Silberholz2007,Tasgetiren2010} and many others.  The advantages of this method are simplicity, compactness (it requires only one integer array of size $m$) and quickness of weight calculation.  The disadvantages are difficulty in some tour modifications (observe that an \insertion{}{} move takes $O(m)$ operations) and absence of a trivial tour correctness test.  In addition, sliding along a tour in this representation requires additional measures to process a tour as a cycle, not as a finite sequence.

Another tour representation, random-key, was used by \citet{Snyder2000}.  It represents the tour as a sequence of real numbers $(x_1, x_2, \ldots, x_m)$; the $i$th number $x_i$ corresponds to the $i$th cluster $C_i$ of the problem.  The integer part $\lfloor x_i \rfloor$ of the number is the vertex index within the cluster $C_i$ and the fractional part $x_i - \lfloor x_i \rfloor$ determines the position of the cluster in the tour---the clusters are ordered according to these fractional parts, in ascending order.  The main advantage of random-key tours is that almost any sequence of numbers represent a correct tour; one only needs to ensure that $1 \le \lfloor x_i \rfloor \le |C_i|$ for every $i$.  It is also relatively easy to implement some modifications of the tour.  The disadvantages are difficulty in sliding along the tour and a high cost of the tour weighing.

We propose a new tour representation which is based on double-linked lists.  We store three integer arrays of size $m$: \var{prev}, \var{next} and \var{vertices}, where $\var{prev}_i$ is a cluster preceding cluster $C_i$ in the tour, $\var{next}_i$ is a cluster succeeding cluster $C_i$ in the tour, and $\var{vertices}_i$ is a vertex within cluster $C_i$.  There are several important advantages of this representation.  Unlike other approaches, it naturally represents the cycle which simplifies the algorithms.  Consider, e.g., a typical local search implementation (Algorithm~\ref{alg:ls_implementation}):
\begin{algorithm}[ht]
\caption{Typical implementation of a local search with a double-linked list based tour representation.  The algorithm performs as few iterations as possible to ensure that the tour is a local minimum.}
\label{alg:ls_implementation}
\begin{algorithmic}
\STATE Initialize current cluster index $i \gets 1$.
\STATE Initialize counter $t \gets m$.
\WHILE {$t > 0$}
	\IF {there exist some improvements for the current cluster $C_i$}
		\STATE Update the tour accordingly.
		\STATE Update the counter $t \gets m$.
	\ELSE
		\STATE Decrease the counter $t \gets t - 1$.
	\ENDIF
	\STATE Move to the next cluster $i \gets \var{next}_i$.
\ENDWHILE
\end{algorithmic}
\end{algorithm}
the algorithm smoothly slides along the tour until no improvement is found for exactly one loop.  Observe that one does not need the concept of position when using this tour representation; it is possible to use cluster index instead.  In this context, the procedure of tour rotation becomes meaningless; one can simply consider any cluster as the first cluster in the tour.  Moreover, it allows one to find a certain cluster in $O(1)$ time; we use it, e.g., to start the \CO{} calculations from the smallest cluster with no extra effort.

Our representation clearly splits the cluster order and the vertex selection; note that some algorithms do not require the information on the vertex selection while some others do not modify the cluster order.  It is useful that linked lists allow quick removing and inserting of elements.  Moreover, to turn the tour backwards, one only needs to swap the arrays \var{prev} and \var{next}.  Observe that this tour representation is deterministic, i.e., each GTSP tour has exactly one representation in this form.  If the problem is symmetric, every tour $(\var{prev}, \var{next}, \var{vertices})$ has exactly one clone $(\var{next}, \var{prev}, \var{vertices})$.  

The main disadvantage of this representation is that it takes three times more space than the sequence of vertices.  In practice, however, many algorithms do not require backward links so one can avoid using the \var{prev} array and reduce the memory usage to two $m$-elements arrays.  When necessary, there is an efficient procedure to restore the \var{prev} array according to \var{next}.

Note that a similar tour representation was used by~\cite{Tasgetiren2007}.

\subsection{Weight Matrix Representation}

Another important decision is how to store the weights in a GTSP instance.  There are two obvious solutions of this problem:
\begin{enumerate}
	\item Store a two dimensional matrix $M$ of size $n \times n$ as follows: $M_{i,j} = w(V_i, V_j)$.  Note that this data structure stores $\sum_{i=1}^m |C_i|^2$ redundant weights.

	\item Store $m (m - 1)$ matrices, one matrix $M^{X,Y}$ of size $|X| \times |Y|$ per every pair of distinct clusters $X$ and $Y$.
\end{enumerate}

If we have a pair of vertices and need to find the weight between them, it is obviously better to use the first approach.  However, if we need to use many weights between two clusters (consider, e.g., calculation of the smallest weight between clusters $X$ and $Y$: $w_\text{min}(X, Y)$), the second approach is preferable.  Indeed, in the first approach we have to look for the absolute index of every vertex in $X$ and $Y$.  In the second approach, we just use the entries of the matrix $M^{X,Y}$.  Observe also that the second approach provides a sequential access to the weight matrix which is friendly with respect to computer architecture and, hence, faster.

Our experimental analysis shows that the second approach improves the performance of \CO{} approximately twice.  However, it is not efficient, e.g., for the Basic adaptations (see Section~\ref{sec:ls_adaptation}).  In our implementations, we store the weights in both forms.

\section{Conclusion}

Three classes of GTSP neighborhoods are selected and discussed in this study.  The most interesting neighborhood in the first class is Cluster Optimization.  Having nice theoretical properties, it can be explored very quickly which makes the \CO{} algorithm an essential subroutine in many heuristics.  Thus, the performance of \CO{} is of great importance.  We introduce several improvements to the algorithm and prove that our implementation almost reaches the best performance possible for this neighborhood.

The TSP-inspired neighborhoods is a large class of neighborhoods derived from TSP neighborhoods.  We formalize the procedure of adaptation of a TSP neighborhood for the GTSP\@.  Among other results, by proposing several new approaches, we significantly speed up, both theoretically and in practice, exploration of the most powerful, `Global', adaptation making it practically useful.  This is particularly interesting since Global adaptation is well-known from the literature and was used or considered many times.  This indicates that there is still great room for further improvements of local search algorithms for GTSP and other fundamental problems.

The neighborhoods of the Fragment Optimization class were not widely used before, probably because of their relatively poor performance.  In this study, we propose an efficient exploration algorithm for the largest neighborhood of this class.  However, this algorithm is not intended to be used as a stand-alone local search.  We believe that it can be very effective as a part of a more sophisticated heuristic.

Further research is required to study possible combinations of GTSP local searches.  We also believe that one can significantly improve the performance of GTSP metaheuristics by using several results of this paper.


{\small
\bibliographystyle{model2-names}
\bibliography{GtspLocalSearch}{}
}

\end{document}